\documentclass[aps,prc,showkeys,nofootinbib]{revtex4}

\usepackage{color} 

\usepackage{graphicx}
\usepackage{dcolumn}
\usepackage{amsthm}

\usepackage{isotope}
\usepackage{braket}
\usepackage{physics}

\usepackage[all]{xy}

\begin{document}

\title{Quantum design in study of pycnonuclear reactions in compact stars and new quasibound states
}
\author{Sergei~P.~Maydanyuk$^{(1,2)}$}
\email{sergei.maydanyuk@wigner.hu}%
\author{Kostiantyn~A.~Shaulskyi$^{(2)}$}\email{konstiger1998@live.com}%

\affiliation{$^{(1)}$Wigner Research Center for Physics, Budapest, 1121, Hungary}
\affiliation{$^{(2)}$Institute for Nuclear Research, National Academy of Sciences of Ukraine, Kyiv, 03680, Ukraine}


\date{\small\today}

\begin{abstract}
Pycnonuclear reactions in the compact stars at zero temperatures are studied on quantum mechanical basis in the paper.
Formalism of multiple internal reflections is generalized for analysis, that was developed for nuclear decays and captures by nuclei with high precision
and tests.
For the chosen reaction $\isotope[12]{C} + \isotope[12]{C} = \isotope[24]{Mg}$ we find the following.
%
%
A quantum study of the pycnonuclear reaction requires a complete analysis of quantum fluxes in the internal nuclear region.
This reduces rate and number of pycnonuclear reactions by 1.8 times.
This leads to the appearance of new states (called as quasibound states) where the compound nuclear system is formed with maximal probability.
%
As shown, minimal energy of such a state is a little higher than energy of zero-point vibrations in lattice sites in pycnonuclear reaction,
however probability of formation of compound system at the quasibound state is essentially larger than
the corresponding probability at state of zero-point vibrations.
%
Hence, there is a sense to tell about reaction rates in such quasibound states as more probable, rather than states of zero-point vibrations.
This can lead to the essential changes in estimation of the rates of nuclear reactions in stars.
%
\end{abstract}


\maketitle

\section{Introduction
\label{sec.introduction}}


In thermonuclear reaction in stars, the thermal energy of the reacting nuclei overcomes the Coulomb repulsion between them so that the reaction can proceed.
At sufficiently high densities, even at zero temperature, zero-point energy of the nuclei in a lattice can lead to an appreciable rate of reactions.
This phenomenon is known as pycnonuclear reaction (from ``pyknos'' as ``dense'' in Greek)~\cite{Cameron.1959b.AstrJ}.
Insight to these reactions was given by Zel'dovich proposed to estimate zero-point energy as some fixed energy of discrete energy spectrum for potential of harmonic oscillatoric type formed at closest approximation at middle point between two nuclei in lattice~\cite{Zeldovich.1965.AstrJ}.
Rates of reactions derived at such zero-point energies are estimated for some atomic nuclei in lattices in compact stars~\cite{ShapiroTeukolsky.2004.book}.

Pycnonuclear burning occurs in dense and cold cores of white dwarfs~\cite{Salpeter_VanHorn.1969.AstrJ} and
in crusts of accreting neutron stars~\cite{Schramm.1990.AstrJ,Haensel.1990.AstronAstrophys}.
Pycnonuclear reactions in compact stars have been studied where stellar matter is considered as multicomponent and dense.
Key process in these reactions is fusion of nuclei with formation of new nucleus with larger mass.
In Ref.~\cite{Yakovlev.2006.PRC} this process was analyzed for reactions containing atomic nuclei of different types,
where authors derived the astrophysical $S$-factors for carbon-oxygen and oxygen-oxygen fusion reactions on the microscopic basis.
Phenomenological expressions for reaction rates containing several fit parameters were found, which can be effective for fast calculations.
In Ref.~\cite{Afanasjev.2010.ADNDT} the astrophysical $S$-factors were estimated for
946 fusion reactions involving stable and neutron-rich isotopes of C, O, Ne, and Mg for center-of-mass energies varying from 2 to $\approx 18$--30~MeV
(see also Ref.~\cite{Singh.2019.NPA}).
Large collection of astrophysical $S$ factors and their compact representation was presented in Ref.~\cite{Afanasjev.2012.PRC} for
isotopes of Be, B, C, N, O, F, Ne, Na, Mg, and Si
(database of $S$ factors was created for about 5000 nonresonant fusion reactions).
Important question in this topic is influence of structure of the multi-component matter (a regular lattice, a uniform mix, etc.) on nuclear processes in reactions which has been studied by many researchers.
%
%
%
But, from previous study it has been known that study of quantum fluxes in the internal region of nuclear system in reaction can essentially change cross-sections and other characteristics of reactions
(see Refs.~\cite{Maydanyuk.2011.JPS,Maydanyuk.2015.NPA,Maydanyuk_Zhang_Zou.2017.PRC}, reference therein).
That question has not been taken into account yet and, so, it is a subject of our current study in this paper.

From previous study of nuclear processes in captures of $\alpha$ particles by nuclei~\cite{Maydanyuk.2015.NPA,Maydanyuk_Zhang_Zou.2017.PRC}
we concluded about importance to take into account internal shape of nuclear potential.
In particular, analysis of quantum processes before fusion shows that resulting cross-sections can be varied up 4 times for energies of beams of $\alpha$ particles used in experiments
(i.e., changes of cross-sections can be larger than 100 percents).
Such changes are controlled by additional independent parameters appeared from fully quantum study.%
\footnote{Simple understanding about new independent parameters can be obtained from comparison of wave function in the first and second approximations in semiclassical approximation.
In particular, wave function in the second approximation includes momentum and force as additional parameters
(for example, see Eq.~(46,9), (46,11) in Ref.~\cite{Landau.v3.1989}, p.~210).
This can be indication to add dynamical characteristics of quantum process (which can be described in the stationary formalism).
More details can be obtained if to study 2D tunneling in quantum mechanics.}
With method in Ref.~\cite{Maydanyuk.2015.NPA}, new parametrization of the $\alpha$-nucleus potential and fusion probabilities were found
(see Fig.~6, Table~2 and B.3 in that paper) 
and error in a description of experimental data is decreased
by $41.72$ times for $\alpha + ^{40}{\rm Ca}$ and $34.06$ times for $\alpha + ^{44}{\rm Ca}$
in comparison with previously existed results (see Fig.~5 and Table~1 in that paper, for details).
Up to present, this is the most accurate approach in description of experimental data for the $\alpha$-capture
(this is calculation given by the blue solid line in Fig.~3~(b) in Ref.~\cite{Maydanyuk_Zhang_Zou.2017.PRC}
for $\alpha + \isotope[44]{Ca}$ in comparison with experimental data~\cite{Eberhard.1979.PRL}).
This pure quantum effect is ignored in estimations of cross-sections based on the semiclassical calculations of penetrability of the barrier.
On such a motivation, we are interesting in analysis of quantum processes for pycnonuclear reactions in compact stars on this fully quantum basis.
In this paper we are focusing on aspects where quantum effects play essential role in reactions of such a type.

Important in that direction is that fully quantum study introduced tests based on quantum mechanics, which allow to check calculations.
In frameworks of approach~\cite{Maydanyuk.2015.NPA,Maydanyuk_Zhang_Zou.2017.PRC},
accuracy of the calculated characteristics is about of $10^{-14}$, while outside such a quantum approach accuracy is based on the semiclassical calculations of the first order, that can be about $10^{-1}$--$10^{-3}$ inside energy region where such an approximation is applied
(if to ignore role of additional independent quantum parameters indicated above).
Moreover, energies for the pycnonuclear reactions are low and there are only processes of deep tunneling under the barrier where the semiclassical approximation cannot be applied at all.
This reinforces our interest to this problem.

The paper is organized in the following way.
In Sec.~\ref{sec.3} a new generalized formalism of multiple internal reflections is presented with focus on description of pycnonuclear reactions.
In Sec.~\ref{sec.analysis} pycnonuclear reaction for $\isotope[12]{C} + \isotope[12]{C} = \isotope[24]{Mg}$ \cite{Gasques.2005.PRC} on the basis of fully quantum approach and without it is given,
supported by calculations of penetrabilities of the barrier, tests, probabilities of formation of the compound nucleus,
estimation of energies and other quantum characteristics for quasibound states,
rates of reactions,
etc..
We summarize conclusions in Sec.~\ref{sec.conclusion}.
In Appendix~\ref{sec.4} formulas for calculations of rates of reactions in compact stars, with fusion probabilities and without those are added.

\section{Tunneling of particle through the barrier on semiaxis (radial problem)
\label{sec.3}}

\subsection{Potential with barrier of the simplest shape:
Multiple internal reflections method in description of scattering of particle (with possibility of capture)
\label{sec.3.1}}

%
We shall study the capture of the particle by the nucleus in the spherically symmetric consideration.
To generalize an idea of the multiple internal reflections on the capture problem and to study packet tunneling
through complicated realistic barriers, let us consider the simplest radial potential, which allows to describe
a scattering of the particle on it with a possible propagation into the internal region.
The simples potential convenient for this aim is%
\footnote{This approximation for barriers of the proton- and $\alpha$-decays for nuclei
with width of each step about 0.01~fm was used previously, stability and convergence of calculations of
amplitudes of the wave function and penetrability were demonstrated (for example, see Ref.~\cite{Maydanyuk.2011.JMP}).
So, we have effective tools for a detailed study of the quantum processes of tunneling and penetrability.}
%
%
\begin{equation}
  V(r) = \left\{
  \begin{array}{cll}
    V_{1}   & \mbox{at } r_{\rm min} < r \leq r_{1}  & \mbox{(region 1)}, \\
    0   & \mbox{at } r_{1} \leq r \leq r_{\rm max}   & \mbox{(region 2)},
  \end{array} \right.
\label{eq.3.1.1}
\end{equation}
%
%
where $V_{1}<0$.
Let us denote the internal and external regions by numbers 1 and 2 (we suppose $r_{\rm min} \ge 0$).
We assume that particle from external region 2 is incident on nucleus with possibility to capture it by the nucleus in the region 1 after
its transfer through a boundary between two regions at $r_{1}$,
or reflection and leaving outside n region 2.
%
A general solution of the wave function has the following form:
\begin{equation}
  \psi(r, \theta, \varphi) =  \frac{\chi(r)}{r} Y_{lm}(\theta, \varphi),
\label{eq.3.1.2}
\end{equation}
\begin{equation}
\chi(r) = \left\{
\begin{array}{lll}
   \alpha_{1}\, e^{ik_{1}r} + \beta_{1}\, e^{-ik_{1}r}  & \mbox{at } r_{\rm min} < r \leq r_{1} & \mbox{(region 1)}, \\
   e^{-ik_{2}r} + A_{R}\,e^{ik_{2}r} & \mbox{at } r_{1} \leq r \leq r_{\rm max} & \mbox{(region 2)},
\end{array} \right.
\label{eq.3.1.3}
\end{equation}
%
%
where $\alpha_{1}$ and $\beta_{1}$ are unknown amplitudes, $A_{T}$ and $A_{R}$ are
unknown amplitudes of transition and reflection, $Y_{lm}(\theta ,\varphi)$ is the spherical function,
and $k_{j} = \frac{1}{\hbar}\sqrt{2m(E-V_{j})}$ are complex wave numbers ($j=1,2$, $V_{2}=0$).
We fix a normalization of the wave function so that a modulus of amplitude of the incident wave $e^{-ik_{2}r}$ equals unity.
We shall search a solution of this problem by the multiple internal reflections approach.

\subsubsection{Approach of step-by-step
\label{sec.3.1.1}}

\begin{figure}[htbp]
$
\begin{xy} 0;<4mm,0mm>:
  ,(0,0)*{}; (10,0)*{}**@{-} 
  ?>*@{>}
  ,(9.8,0.5)*\txt{$r$}
  ,(0,-3.5)*{}; (0,5)*{}**@{-}
  ?>*@{>}
  ,(1.2,4.5)*\txt{$V(r)$}
  ,(7.7,-2.7)*\txt{(a)}
  ,(0,-3.0)*{}; (5,-3.0)*{}**@{-} 
  ,(5,-3.0)*{}; (5,0)*{}**@{-}
  ,(5,0)*{}; (5,4.5)*{}**@{--}
  ,(5.7,-0.6)*\txt{$r_{1}$}
  ,(6,3)*{}; (8.0,3)*{}**@{.} 
  ?<*@{<}
  ?(0.5)*!/_3.0mm/{\txt{\small{$\chi_{\rm inc}^{(1)}$} }}
  ,(2,2.25)*{}; (4,2.25)*{}**@{.} 
  ?<*@{<}
  ?(0.5)*!/_3.0mm/{\txt{\small{$\chi_{\rm tr}^{(1)}$} }}
  ,(6,1.2)*{}; (8.0,1.2)*{}**@{.} 
  ?(0.5)*!/_3.0mm/{\txt{\small{$\chi_{\rm ref}^{(1)}$} }}
  ?>*@{>}
\end{xy}
$
\hspace{5mm}
\begin{xy} 0;<4mm,0mm>:
  ,(0,0)*{}; (10,0)*{}**@{-} 
  ?>*@{>}
  ,(9.8,0.5)*\txt{$r$}
  ,(0,-3.5)*{}; (0,5)*{}**@{-}
  ?>*@{>}
  ,(1.2,4.5)*\txt{$V(r)$}
  ,(7.7,-2.7)*\txt{(b)}
  ,(0,-3.0)*{}; (5,-3.0)*{}**@{-} 
  ,(5,-3.0)*{}; (5,0)*{}**@{-}
  ,(5,0)*{}; (5,4.5)*{}**@{--}
  ,(5.7,-0.6)*\txt{$r_{1}$}
  ,(1.0,3.0)*{}; (3.0,3.0)*{}**@{.} 
  ?<*@{<}
  ?(0.5)*!/_3.0mm/{\txt{\small{$\chi_{\rm inc}^{(2)}$} }}
  ,(1.0,1.2)*{}; (3.0,1.2)*{}**@{.} 
  ?(0.5)*!/_3.0mm/{\txt{\small{$\chi_{\rm ref}^{(2)}$} }}
  ?>*@{>}
\end{xy}
\caption{\small 
The incident, transmitted and reflected waves in the first step (a), the second step (b).
\label{fig.2.1}}
\end{figure}
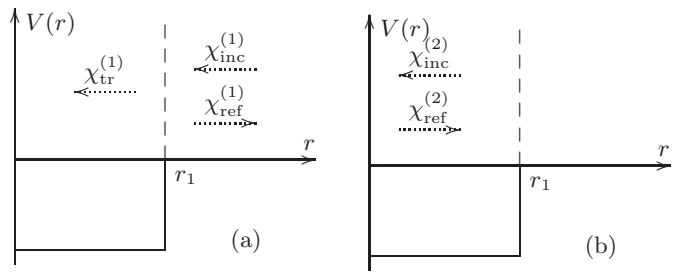

%
%
According to the multiple internal reflections method,
the scattering of a particle on the barrier is sequentially considered by steps of propagation of the wave (wave packet)
relative to each boundary of the barrier
(idea of this approach can be understood most clearly in the problem of tunneling through the simplest
rectangular barrier,
see Refs.~\cite{
Maydanyuk.2000.UPJ,Maydanyuk.2002.JPS,Maydanyuk.2006.FPL},
%
where one can find proof of this fully quantum exactly solvable method and analyze its properties in details).
In the first step we consider the wave $e^{-ik_{2}r}$ in region 2, which is incident on the boundary at $r_{1}$  outside.
This wave is transformed into new wave $\beta_{1}^{(1)} e^{-ik_{1}r}$ propagated to the center in the region 1, and
new wave $\alpha_{2}^{(1)} e^{ik_{2}r}$ which is reflected from the boundary and propagated outside in the region 2
[see Fig.~\ref{fig.2.1}~(a)].
%
For such a process we have such wave function:
\begin{equation}
\chi^{(1)}(r) = \left\{
\begin{array}{lll}
   \beta_{1}^{(1)}\, e^{-ik_{1}r}  & \mbox{at } r_{\rm min} < r \leq r_{1}, \\
   e^{-ik_{2}r} + \alpha_{2}^{(1)}\,e^{ik_{2}r} & \mbox{at } r_{1} \leq r \leq r_{\rm max}.
\end{array} \right.
\label{eq.3.1.1.1}
\end{equation}
%
%
Here, $\alpha_{2}^{(1)}$ and $\beta_{1}^{(1)}$ are two unknown amplitudes (we add upper index denoting number of step).
We find them from condition of continuity of the full wave function and its derivative at $r_{1}$:
\begin{equation}
\begin{array}{ll}
  \alpha_{2}^{(1)} =
    R_{1}^{-} =
    \displaystyle\frac{k-k_{1}}{k+k_{1}}\, e^{-2ikr_{1}}, &
  \beta_{1}^{(1)} =
    T_{1}^{-} =
    \displaystyle\frac{2k}{k+k_{1}}\, e^{-i(k-k_{1})r_{1}},
\end{array}
\label{eq.3.1.1.2}
\end{equation}
%
where
\begin{equation}
\begin{array}{ll}
  R_{1}^{-} = \displaystyle\frac{k-k_{1}}{k+k_{1}}\, e^{-2ikr_{1}}, &
  T_{1}^{-} = \displaystyle\frac{2k}{k+k_{1}}\, e^{-i(k-k_{1})r_{1}}.
\end{array}
\label{eq.3.1.1.3}
\end{equation}
%
%
Now, let us analyze a sense of these new waves.
The transmitted wave is formed inside the internal nuclear region.
So, it describes formation of the compound nuclear system and its further evolution.
The reflected wave does not reach space region of the nucleus, it describes reflection outside by Coulomb forces of nucleus.
So, it can characterize the potential scattering, without a stage with formation of the compound nuclear system.
So, on the basis of this extremely simple scheme, we separate two physically different processes,
i.e. (1) the resonant scattering with stage of formation of the compound nuclear system and
(2) the potential scattering without the compound nucleus formation.
%
%
For convenience, we have introduced two new amplitudes $T_{1}^{-}$ and $R_{1}^{+}$, describing propagation and reflection concerning to the studied boundary (bottom index ``1' ' indicates number of boundary, upper index ``$-$'' or ``$+$'' indicates type of the amplitude concerning to the incident flux, which is directed in the negative or positive radial direction, respectively).

%
On the second step we consider the wave $\beta_{1}^{(1)}\, e^{-ik_{1}r}$ transmitted into the region 1 through
the boundary in the previous step.
This wave moves to center and is transformed into a new wave which moves in positive direction in this region.
We have such a wave function for this process:
\begin{equation}
\begin{array}{ll}
  \chi^{(2)}(r) =
  \beta_{1}^{(1)}\, e^{-ik_{1}r} + \alpha_{1}^{(2)}\,e^{ik_{1}r} & \mbox{at } r_{\rm min} < r \leq r_{1}.
\end{array}
\label{eq.3.1.1.4}
\end{equation}
%
Here, $\alpha_{1}^{(2)}$ is a new unknown amplitude, which is found from a condition of continuity of the wave function at boundary $r_{\rm min}$.
At $r_{\rm min}=0$ we have:
\begin{equation}
\begin{array}{ll}
  \alpha_{1}^{(2)} = R_{0}\, \beta_{1}^{(1)}, & R_{0} = -1.
\end{array}
\label{eq.3.1.1.5}
\end{equation}
%
But, at $r_{\rm min} \ge 0$ we have
\begin{equation}
\begin{array}{ll}
  \alpha_{1}^{(2)} = R_{0}\, \beta_{1}^{(1)}, & R_{0} = - e^{-2ik_{1}r_{\rm min}}.
\end{array}
\label{eq.3.1.1.6}
\end{equation}

%
On the third step we consider the wave $\alpha_{1}^{(2)}\,e^{ik_{1}r}$ which was formed in the previous step
and is propagated outside in the region 1.
This wave is incident on the boundary at point $r_{1}$ and is transformed into
a new wave which transmitted through the boundary and is moving outside in positive direction in region 2,
and a new reflected wave in region 1.
We have such a wave function for this process:
\begin{equation}
\chi^{(3)}(r) = \left\{
\begin{array}{lll}
   \alpha_{1}^{(2)}\,e^{ik_{1}r} + \beta_{1}^{(3)}\, e^{-ik_{1}r}  & \mbox{at } r_{\rm min} < r \leq r_{1}, \\
   \alpha_{2}^{(3)}\,e^{ik_{2}r} & \mbox{at } r_{1} \leq r \leq r_{\rm max}.
\end{array} \right.
\label{eq.3.1.1.7}
\end{equation}
%
%
Here, $\alpha_{2}^{(3)}$ and $\beta_{1}^{(3)}$ are two unknown amplitudes.
These amplitudes are found from condition of continuity of the full wave function and its derivative at $r_{1}$:
\begin{equation}
\begin{array}{ll}
  \alpha_{2}^{(3)} =
    \alpha_{1}^{(2)}\, T_{1}^{+} =
    \alpha_{1}^{(2)}\, \displaystyle\frac{2k_{1}}{k+k_{1}}\, e^{i(k_{1}-k)r_{1}}, &
  \beta_{1}^{(3)} =
    \alpha_{1}^{(2)}\, R_{1}^{+} =
    \alpha_{1}^{(2)}\, \displaystyle\frac{k_{1}-k}{k+k_{1}}\, e^{2ik_{1}r_{1}},
\end{array}
\label{eq.3.1.1.8}
\end{equation}
%
where
\begin{equation}
\begin{array}{ll}
  T_{1}^{+} = \displaystyle\frac{2k_{1}}{k+k_{1}}\, e^{i(k_{1}-k)r_{1}}, &
  R_{1}^{+} = \displaystyle\frac{k_{1}-k}{k+k_{1}}\, e^{2ik_{1}r_{1}}.
\end{array}
\label{eq.3.1.1.9}
\end{equation}


Each step in such a consideration of propagation of waves is similar to one of the first independent $2N-1$ steps.
From the analysis of these steps we obtain recurrent relations for the calculation of the unknown amplitudes
$\alpha_{1}^{(n)}$, $\beta_{1}^{(n)}$ and $\alpha_{2}^{(n)}$ for an arbitrary step with number $n$
(here, 
the logic of the definition of these amplitudes can be found in Appendix~A in~\cite{Maydanyuk.2011.JMP}).
In the summation of these relations from each step,
we impose the continuity condition for the full (summarized) wave function and its derivative relative to the corresponding boundary.


\subsubsection{Summations of amplitudes
\label{sec.3.1.3}}

We calculate summations of all amplitudes of all waves in region 1:
\begin{equation}
\begin{array}{ll}
  \displaystyle\sum\limits_{i=1} \beta_{1}^{(i)} =
  T_{1}^{-}\, \Bigl( 1 + \displaystyle\sum\limits_{i=1} (R_{0} R_{1}^{+})^{i} \Bigr) =
  \displaystyle\frac{T_{1}^{-}}{1 - R_{0} R_{1}^{+}}, \\

  \displaystyle\sum\limits_{i=1} \alpha_{1}^{(i)} =
  R_{0}\, \displaystyle\sum\limits_{i=1} \beta_{1}^{(i)} =
  \displaystyle\frac{R_{0} T_{1}^{-}}{1 - R_{0} R_{1}^{+}}.
\end{array}
\label{eq.3.1.3.1}
\end{equation}
%
Summation of all amplitudes of all outgoing waves in the region 2 outside is
\begin{equation}
\begin{array}{ll}
  \displaystyle\sum\limits_{i=1} \alpha_{2}^{(i)} = \alpha_{2}^{(1)} + \displaystyle\sum\limits_{i=2} \alpha_{2}^{(i)}, \\
  \displaystyle\sum\limits_{i=2} \alpha_{2}^{(i)} =
  T_{1}^{-} R_{0} T_{1}^{+} \Bigl( 1 + \displaystyle\sum\limits_{i=1} (R_{0} R_{1}^{+})^{i} \Bigr) =
  \displaystyle\frac{T_{1}^{-} R_{0} T_{1}^{+} }{1 - R_{0} R_{1}^{+}}.
\end{array}
\label{eq.3.1.3.2}
\end{equation}
%
We define the following characteristic
\begin{equation}
  A_{\rm osc} = \displaystyle\frac{1}{1 - R_{0} R_{1}^{+}},
\label{eq.3.1.3.3}
\end{equation}
%
and we rewrite summations as
\begin{equation}
\begin{array}{lll}
  \displaystyle\sum\limits_{i=1} \beta_{1}^{(i)} = A_{\rm osc}\, T_{1}^{-}, &

  \displaystyle\sum\limits_{i=1} \alpha_{1}^{(i)} =
  R_{0}\, \displaystyle\sum\limits_{i=1} \beta_{1}^{(i)} =
  A_{\rm osc}\, R_{0} T_{1}^{-}, &

  \displaystyle\sum\limits_{i=2} \alpha_{2}^{(i)} = A_{\rm osc}\, T_{1}^{-} R_{0} T_{1}^{+}.
\end{array}
\label{eq.3.1.3.4}
\end{equation}
%
This characteristic describes oscillations of waves inside nuclear region up to barrier maximum
(so we call $A_{\rm osc}$ as amplitude of oscillations).
We calculate $A_{\rm osc}$ (at $R_{0}=-1$):
\begin{equation}
  A_{\rm osc} =
  \displaystyle\frac{k + k_{1}}{(k+k_{1}) + (k_{1}-k)\,e^{i2k_{1}r_{1}}} =
  \displaystyle\frac{(k + k_{1})\, [(k+k_{1}) + (k_{1}-k)\,e^{-i2k_{1}r_{1}}]}
    {2k^{2} (1 -\cos (2k_{1}r_{1})) + 2k_{1}^{2}\,(1 + \cos (2k_{1}r_{1})) }.
\label{eq.3.1.3.5}
\end{equation}
%
Now we calculate summations of other amplitudes (at $R_{0}=-1$):
\begin{equation}
\begin{array}{lclll}
\vspace{2mm}
  \displaystyle\sum\limits_{i=1} \beta_{1}^{(i)} & = &
  A_{\rm osc} \cdot T_{1}^{-} =
  \displaystyle\frac{2k\, e^{-i(k-k_{1})r_{1}}} {(k+k_{1}) + (k_{1}-k)\,e^{i2k_{1}r_{1}}}, \\

\vspace{2mm}
  \displaystyle\sum\limits_{i=1} \alpha_{1}^{(i)} & = &
    R_{0} \cdot \displaystyle\sum\limits_{i=1} \beta_{1}^{(i)} =
    - \displaystyle\frac{2k\, e^{-i(k-k_{1})r_{1}}} {(k+k_{1}) + (k_{1}-k)\,e^{i2k_{1}r_{1}}}, \\

  \displaystyle\sum\limits_{i=2} \alpha_{2}^{(i)} & = &
  A_{\rm osc} \cdot T_{1}^{-}R_{0}T_{1}^{+} =
%
    -\, \displaystyle\frac{4kk_{1}}{(k+k_{1}) + (k_{1}-k)\,e^{i2k_{1}r_{1}}}\,
    \displaystyle\frac{1}{k+k_{1}}\, e^{2i(k_{1}-k)r_{1}}.
\end{array}
\label{eq.3.1.3.6}
\end{equation}

Now we calculated square of modulus of this amplitude $A_{\rm osc}$ [from Eq.~(\ref{eq.3.1.3.5})]
\begin{equation}
\begin{array}{lllll}
\vspace{2mm}
  |A_{\rm osc}|^{2} & = &
%
  \displaystyle\frac{(k + k_{1})^{2}} {2k^{2} (1 -\cos (2k_{1}r_{1})) + 2k_{1}^{2}\,(1 + \cos (2k_{1}r_{1})) },
\end{array}
\label{eq.3.1.3.7}
\end{equation}
%
modulus of this amplitude
\begin{equation}
\begin{array}{lllll}
\vspace{2mm}
  |A_{\rm osc}| & = &
  \displaystyle\frac{k + k_{1}}
    {\sqrt{2k^{2} (1 -\cos (2k_{1}r_{1})) + 2k_{1}^{2}\,(1 + \cos (2k_{1}r_{1})) }}.
\end{array}
\label{eq.3.1.3.8}
\end{equation}
%
Also we calculate summations of modulus of all amplitudes in the region 1:
\begin{equation}
\begin{array}{lll}
  |\displaystyle\sum\limits_{i=1} \beta_{1}^{(i)}|^{2} =
  |\displaystyle\sum\limits_{i=1} \alpha_{1}^{(i)}|^{2} =
  \displaystyle\frac{2k^{2}} {k^{2} (1 -\cos (2k_{1}r_{1})) + k_{1}^{2}\,(1 + \cos (2k_{1}r_{1}))},
\end{array}
\label{eq.3.1.3.9}
\end{equation}
%
and summations of amplitudes of all outgoing waves in region 2:
\begin{equation}
\begin{array}{lll}
  |\alpha_{2}^{(1)}|^{2} & = &
  |R_{1}^{-}|^{2} =
  \displaystyle\frac{(k - k_{1})^{2}} {(k + k_{1})^{2}}, \\

  |\displaystyle\sum\limits_{i=2} \alpha_{2}^{(i)}|^{2} & = &
  |A_{\rm osc}|^{2}\, |T_{1}^{-} R_{0} T_{1}^{+}|^{2} =

%
  \displaystyle\frac{8k^{2}k_{1}^{2}}
    {(k + k_{1})^{2}\, [k^{2} (1 -\cos (2k_{1}r_{1})) + k_{1}^{2}\,(1 + \cos (2k_{1}r_{1})) ]}.
\end{array}
\label{eq.3.1.3.10}
\end{equation}
%
We rewrite the final results:
\begin{equation}
\begin{array}{lll}
\vspace{1.5mm}
  \displaystyle\sum\limits_{i=1} \beta_{1}^{(i)} & = &
  -\, \displaystyle\sum\limits_{i=1} \alpha_{1}^{(i)} =
  A_{\rm osc} \cdot T_{1}^{-} =
  \displaystyle\frac{2k\, e^{-i(k-k_{1})r_{1}}} {(k+k_{1}) + (k_{1}-k)\,e^{i2k_{1}r_{1}}}, \\

\vspace{2mm}
  \displaystyle\sum\limits_{i=2} \alpha_{2}^{(i)} & = &
    A_{\rm osc} \cdot T_{1}^{-}R_{0}T_{1}^{+} =
    -\, \displaystyle\frac{4kk_{1}}{(k+k_{1}) + (k_{1}-k)\,e^{i2k_{1}r_{1}}}\,
    \displaystyle\frac{1}{k+k_{1}}\, e^{2i(k_{1}-k)r_{1}}, \\

\vspace{2mm}
  A_{\rm osc} & = &
  \displaystyle\frac{k + k_{1}}{(k+k_{1}) + (k_{1}-k)\,e^{i2k_{1}r_{1}}} =
  \displaystyle\frac{(k + k_{1})\, [(k+k_{1}) + (k_{1}-k)\,e^{-i2k_{1}r_{1}}]}
    {2k^{2} (1 -\cos (2k_{1}r_{1})) + 2k_{1}^{2}\,(1 + \cos (2k_{1}r_{1})) }, \\

  |A_{\rm osc}| & = &
  \displaystyle\frac{k + k_{1}}
    {\sqrt{2k^{2} (1 -\cos (2k_{1}r_{1})) + 2k_{1}^{2}\,(1 + \cos (2k_{1}r_{1})) }},
\end{array}
\label{eq.3.1.3.11}
\end{equation}
%
%
\begin{equation}
\begin{array}{lll}
\vspace{1.5mm}
  |A_{\rm osc}|^{2} & = &
  \displaystyle\frac{(k + k_{1})^{2}}
    {2k^{2} (1 -\cos (2k_{1}r_{1})) + 2k_{1}^{2}\,(1 + \cos (2k_{1}r_{1})) }, \\

  |\displaystyle\sum\limits_{i=1} \beta_{1}^{(i)}|^{2} & = &
  |\displaystyle\sum\limits_{i=1} \alpha_{1}^{(i)}|^{2} =
  |A_{\rm osc}|^{2}\, |T_{1}^{-}|^{2} =
  \displaystyle\frac{2k^{2}} {k^{2} (1 -\cos (2k_{1}r_{1})) + k_{1}^{2}\,(1 + \cos (2k_{1}r_{1}))}, \\

  |\alpha_{2}^{(1)}|^{2} & = &
  |R_{1}^{-}|^{2} =
  \displaystyle\frac{(k - k_{1})^{2}} {(k + k_{1})^{2}}, \\

  |\displaystyle\sum\limits_{i=2} \alpha_{2}^{(i)}|^{2} & = &
  |A_{\rm osc}|^{2}\, |T_{1}^{-} R_{0} T_{1}^{+}|^{2} =
  \displaystyle\frac{8k^{2}k_{1}^{2}}
    {(k + k_{1})^{2}\, [k^{2} (1 -\cos (2k_{1}r_{1})) + k_{1}^{2}\,(1 + \cos (2k_{1}r_{1})) ]}.
\end{array}
\label{eq.3.1.3.12}
\end{equation}


From (\ref{eq.3.1.3.12}) it follows that $A_{\rm osc}$ has no zero-points and divergencies.
We calculate
\begin{equation}
\begin{array}{lllll}
\vspace{1.5mm}
  \displaystyle\frac{d |A_{\rm osc}|^{2}}{dE} & = &
  \displaystyle\frac{m}{\hbar^{2}kk_{1}}\,
      \displaystyle\frac{(k + k_{1})^{2} (k_{1} - k)}
        {[k^{2} (1 -\cos (2k_{1}r_{1})) + k_{1}^{2}\,(1 + \cos (2k_{1}r_{1})) ]^{2}}\; \times \\
  & \times &
    \Bigl\{
      (k_{1}-k) +
      (k_{1} + k) \cos (2k_{1}r_{1}) +
      (k + k_{1}) \sin (2k_{1}r_{1})\, kr_{1}
    \Bigr\}
\end{array}
\label{eq.3.1.4.5}
\end{equation}
and obtain condition for extremums of the amplitude $A_{\rm osc} (E)$ (at $k_{1} \ne k$)
\begin{equation}
  k_{1}-k + (k_{1} + k) \cos(2k_{1}r_{1}) + kr_{1}\,(k_{1} + k) \sin(2k_{1}r_{1}) = 0.
\label{eq.3.1.4.6}
\end{equation}
Note that the incident wave is normalized on unity, but amplitude $A_{\rm osc}$ can be larger than unity.
According to Eq.~(\ref{eq.3.1.3.12}), $A_{\rm osc}$ can be closer to maximums at $\cos(2k_{1}r_{1})= \pm 1$:
\begin{equation}
\begin{array}{lllll}
\vspace{2mm}
  \Bigl\{ \cos(2k_{1}r_{1})= + 1 \Bigr\} \quad \to \quad
  \Bigl\{
    |A_{\rm osc}|^{2} =
    \displaystyle\frac{(k + k_{1})^{2}}{4k_{1}^{2} } =
    \displaystyle\frac{1}{4}\, \Bigl( 1 + \displaystyle\frac{k}{k_{1} } \Bigr)^{2}
    \le 1
  \Bigr\}, \\

  \Bigl\{ \cos(2k_{1}r_{1})= - 1 \Bigr\} \quad \to \quad
  \Bigl\{
    |A_{\rm osc}|^{2} =
    \displaystyle\frac{(k + k_{1})^{2}}{4k^{2} } =
    \displaystyle\frac{1}{4}\, \Bigl( 1 + \displaystyle\frac{k_{1}}{k} \Bigr)^{2}
    \ge 1
  \Bigr\}.
\end{array}
\label{eq.3.1.4.2}
\end{equation}
Energies for the second maximums are (at $n \in \cal{N}$)
\begin{equation}
  E_{\rm max,\, n} = \displaystyle\frac{\hbar^{2}\pi^{2}\, (1 + 2n)^{2}}{8mr_{1}^{2}} - |V_{1}|.
\label{eq.3.1.4.3}
\end{equation}

\subsubsection{
Probability of existence of the compound nuclear system (without fusion)
\label{sec.3.1.5}}

Let us find integral from square of modulus of wave function inside internal region:
%
\begin{equation}
\begin{array}{lll}
\vspace{0.3mm}
  \displaystyle\int\limits_{0}^{r_{1}} |\chi(r)|^{2}\; dr & = &

  \displaystyle\int\limits_{0}^{r_{1}}
    \Bigl|
      \displaystyle\sum\limits_{i=1}
        \alpha_{1}^{(i)} e^{ik_{1}r} +
      \displaystyle\sum\limits_{i=1}
        \beta_{1}^{(i)} e^{-ik_{1}r} \Bigr|^{2} \;dr =

%

  \Bigl| \displaystyle\sum\limits_{i=1} \beta_{1}^{(i)} \Bigr|^{2}\,
  \displaystyle\int\limits_{0}^{r_{1}}
    \Bigl| R_{0}\, e^{ik_{1}r} + e^{-ik_{1}r} \Bigr|^{2} \;dr.
\end{array}
\label{eq.3.1.5.1}
\end{equation}
At $R_{0}=-1$ we have
%
\begin{equation}
\begin{array}{lllll}
  \displaystyle\int\limits_{0}^{r_{1}} |\chi(r)|^{2}\;dr =


  2\, \Bigl| \displaystyle\sum\limits_{i=1} \beta_{1}^{(i)} \Bigr|^{2}\, \Bigl( r_{1} - \displaystyle\frac{\sin(2k_{1}r_{1})}{2k_{1}} \Bigr).
\end{array}
\label{eq.3.1.5.2}
\end{equation}
Taking summation $\sum\limits_{i=1} \beta_{1}^{(i)}$ in Eq.~(\ref{eq.3.1.3.10}) into account, we write
%
\begin{equation}
\begin{array}{lll}
  \displaystyle\int\limits_{0}^{r_{1}} |\chi(r)|^{2}\;dr =
  2\, |A_{\rm osc}|^{2}\, |T_{1}^{-}|^{2}
    \Bigl( r_{1} - \displaystyle\frac{\sin(2k_{1}r_{1})}{2k_{1}} \Bigr).
\end{array}
\label{eq.3.1.5.3}
\end{equation}
%
%
Let us understand physical meaning of the obtained formula. The integral of the square of the wave function over nuclear region determines a probability of existence of a compound nuclear system (within the spatial region from 0 to $r_{1}$), which can be formed during the scattering of particle on nucleus (denoted as $P_{ \rm cn}$). One can see that this probability varies with the energy of the incident particle, and reaches maximums at certain values of energy.
The amplitude squared of $T_{1}^{-}$ is the penetrability coefficient $T_{\rm bar}$, related via an additional factor $k_{1}/k_{2}$.
On this basis, we rewrite the formula above as
\begin{equation}
\begin{array}{llllll}
  P_{\rm cn}^{\rm (without\, fusion)} =
  \displaystyle\int\limits_{0}^{r_{1}} |\chi(r)|^{2}\; dr =
  P_{\rm osc}\, T_{\rm bar}\, P_{\rm loc}, &
  P_{\rm osc} = |A_{\rm osc}|^{2}, &
  T_{\rm bar} \equiv \displaystyle\frac{k_{1}}{k_{2}}\; \bigl| T_{1}^{-} \bigr|^{2}, &
  P_{\rm loc} = 2\, \displaystyle\frac{k_{2}}{k_{1}}\; \Bigl( r_{1} - \displaystyle\frac{\sin(2k_{1}r_{1})}{2k_{1}} \Bigr).
\end{array}
\label{eq.3.1.5.4}
\end{equation}
%
%
So, we have obtained a generalization of idea of Gamow, which is implemented in many papers in calculations of capture cross sections (as well as in similar calculations of half-lives in nuclear decays).
However, in addition to the coefficient of penetrability and the factor describing internal oscillations in the spatial region of the nucleus, we have a new coefficient $P_{\rm loc}$.
It can be associated with the spatial distribution of the wave function of the particle inside the nucleus (it is not related with internal oscillations and the reflectivity of the barrier).
In other words, this is a completely new characteristic (and the process corresponding to it), which was not taken into account by Gamow and his followers in determination of half-life for $\alpha$-decay.
We call it as \emph{coefficient of localization}.

\subsubsection{Test via fluxes
\label{sec.3.1.6}}

Now we calculate square of modulus of the full summation of all amplitudes $\alpha_{2}^{(i)}$ and find:
%
\begin{equation}
  \Bigl|\, \alpha_{2}^{(1)} + \displaystyle\sum\limits_{i=2} \alpha_{2}^{(i)}) \Bigr|^{2} = 1.
\label{eq.3.1.6.1}
\end{equation}
%
%
We conclude that full outgoing flux equals to incident flux from outside in the first step, but it is with opposite direction.
We find an interference term between the reflected wave in the first step (describing the potential
scattering without stage of the compound nucleus formation)
and summation of all other waves propagating outside (which is connected with formation of the compound nucleus):
\begin{equation}
\begin{array}{lll}
  P_{\rm interf} \equiv
  2\; | \alpha_{2}^{(1)*} \cdot \displaystyle\sum\limits_{i=2} \alpha_{2}^{(i)})| & = &
  \displaystyle\frac{4\sqrt{2}\,kk_{1}\, |k-k_{1}|}{(k+k_{1})^{2}}\,
  \displaystyle\frac{1}{\sqrt{ k^{2} (1 -\cos (2k_{1}r_{1})) + k_{1}^{2}\,(1 + \cos (2k_{1}r_{1})} }.
\end{array}
\label{eq.3.1.6.2}
\end{equation}
%
Let us write this term in a different form, showing explicitly the amplitude of the oscillations.
Taking Eq.~(\ref{eq.3.1.3.11}) into account, we have ($R_{0} = -1$):
\begin{equation}
\begin{array}{lll}
  P_{\rm interf} & = &
  2\; | R_{1}^{-} \cdot A_{\rm osc} \cdot R_{0}\, T_{1}^{-}\, T_{1}^{+}| =
%
  | A_{\rm osc}| \cdot \displaystyle\frac{8kk_{1}\,|k-k_{1}|}{(k+k_{1})^{3}}.
\end{array}
\label{eq.3.1.6.3}
\end{equation}
%
One can see that the interference term is not a constant, but a function depending on energy of the incident wave.
This term has maxima and minima at energies close to (but not equal to) the energies of the maxima and minima of the oscillation amplitude $A_{\rm osc}$.
The probability of formation of a compound nucleus can be analyzed in terms of this interference term.

\subsubsection{Fusion in the internal region
\label{sec.3.1.7}}

%
Let us understand how to include fusion in the processes studied above.
From mathematical point of view, one can describe the fusion of an particle with a nucleus in the compound nuclear system,
suppressing waves in the internal region of the nuclear system (i.e., region 1).
Let us consider the full amplitudes (\ref{eq.3.1.3.1}) in this region.
%
%
%
%
The complete fusion can be described if to suppress fluxes for each wave in the internal region of the nucleus to zero.
This can be realized via the following condition of \emph{maximally fast fusion}:
\begin{equation}
  R_{0} \to 0,
\label{eq.3.1.7.2}
\end{equation}
%
and from Eq.~(\ref{eq.3.1.3.1}) we obtain
\begin{equation}
\begin{array}{ll}
  \displaystyle\sum\limits_{i=1} \beta_{1}^{(i)} = T_{1}^{-}, &
  \displaystyle\sum\limits_{i=1} \alpha_{1}^{(i)} = 0.
\end{array}
\label{eq.3.1.7.3}
\end{equation}
%
%
%
Let us calculate summations of the amplitudes going outside into region 2:
\begin{equation}
\begin{array}{ll}
  \displaystyle\sum\limits_{i=2} \alpha_{2}^{(i)} =
  \displaystyle\frac{T_{1}^{-} R_{0} T_{1}^{+} }{1 - R_{0} R_{1}^{+}} = 0, &
  \displaystyle\sum\limits_{i=1} \alpha_{2}^{(i)} =
  \alpha_{2}^{(1)} =
  \displaystyle\frac{k-k_{1}}{k+k_{1}}\, e^{-2ikr_{1}},
\end{array}
\label{eq.3.1.7.4}
\end{equation}
%
and amplitude of oscillations
\begin{equation}
  A_{\rm osc} = \displaystyle\frac{1}{1 - R_{0} R_{1}^{+}} = 1.
\label{eq.3.1.7.5}
\end{equation}
%
%
In this case there is potential scattering but no resonant scattering.
From Eq.~(\ref{eq.3.1.5.1}) and (\ref{eq.3.1.7.2}), the probability of the existence of a compound nucleus is
\begin{equation}
\begin{array}{lll}
\vspace{0.5mm}
  P_{\rm cn}^{\rm (fast\, fusion)} & = &

  \Bigl| \displaystyle\sum\limits_{i=1} \beta_{1}^{(i)} \Bigr|^{2}\,
  \displaystyle\int\limits_{0}^{r_{1}}
    \Bigl| R_{0}\, e^{ik_{1}r} + e^{-ik_{1}r} \Bigr|^{2} \;dr =



  \bigl| T_{1}^{-} \bigr|^{2}\, r_{1} =
  \displaystyle\frac{k_{2}\, r_{1}}{k_{1}}\, T_{\rm bar}.
\end{array}
\label{eq.3.1.7.6}
\end{equation}

%
If we want to construct a compound nucleus existed for some period of time without way of fast fusion,
we should not suppress completely fluxes in the nuclear region, i.e. it needs to use condition less strict than Eq.~(\ref{eq.3.1.7.2}).
We introduce a new factor of fusion $p_{1}$, redefining the reflection amplitude $R_{0}$ as
\begin{equation}
\begin{array}{lll}
  R_{0}^{\rm (old)} \to R_{0} \cdot (1 - p_{1}), &
  0 \le p_{1} \le 1.
\end{array}
\label{eq.3.1.7.7}
\end{equation}
%
%
For $p_{1}=1$ the formula (\ref{eq.3.1.7.7}) turns into (\ref{eq.3.1.7.2}) and we get the fast fusion,
while for $p_{1}=0$ we have the compound nuclear system, which decays (as in the elastic scattering).
%
%
%

\subsection{Potential with barrier of arbitrary shape:
Method of multiple internal reflections in description of collision of nuclei (with possibility of fusion)
\label{sec.3.2}}

%
Let us consider a process of capture of the particle by the nucleus with the radial barrier of arbitrary shape,
which has successfully been approximated by a sufficiently large number $N$ of rectangular steps:
\begin{equation}
  V(r) = \left\{
  \begin{array}{cll}
    V_{1}   & \mbox{at } r_{\rm min} < r \leq r_{1}      & \mbox{(region 1)}, \\
    \ldots   & \ldots & \ldots \\
    V_{N_{\rm cap}}   & \mbox{at } r_{N_{\rm cap} - 1} \leq r \leq r_{\rm cap}         & \mbox{(region $N_{\rm cap}$)}, \\
    \ldots   & \ldots & \ldots \\
    V_{N}   & \mbox{at } r_{N-1} \leq r \leq r_{\rm max} & \mbox{(region $N$)},
  \end{array} \right.
\label{eq.3.2.1.1}
\end{equation}
%
%
%
%
where $V_{j}$ are constants ($j = 1 \ldots N$). %
As in the previous section, we denote the first region with a left boundary at point $r_{\rm min}$. (we assume $r_{\rm min} \le 0$).
In addition to study in Ref.~\cite{Maydanyuk.2015.NPA},
now we shall assume that the capture of the particle by the nucleus takes place the most probably in region with number $N_{\rm capture}$
after its tunneling through the barrier,
and further propagation of waves to inside internal region exists and, so, should be studied also.
This logic follows from condition of continuity of fluxes in quantum mechanics, that is strict condition, and it will better to include it to analysis also.
A general solution of the radial wave function (up to its normalization) for the above barrier energies has the following form:
\begin{equation}
\chi(r) = \left\{
\begin{array}{lll}
   \alpha_{1}\, e^{ik_{1}r} + \beta_{1}\, e^{-ik_{1}r},
     & \mbox{at } r_{\rm min} < r \leq r_{1} & \mbox{(region 1)}, \\
   \alpha_{2}\, e^{ik_{2}r} + \beta_{2}\, e^{-ik_{2}r},
     & \mbox{at } r_{1} \leq r \leq r_{2} & \mbox{(region 2)}, \\
   \ldots & \ldots & \ldots \\
   \alpha_{N-1}\, e^{ik_{N-1}r} + \beta_{N-1}\, e^{-ik_{N-1}r}, &
     \mbox{at } r_{N-2} \leq r \leq r_{N-1} & \mbox{(region $N-1$)}, \\
   e^{-ik_{N}r} + A_{R}\,e^{ik_{N}r}, & \mbox{at } r_{N-1} \leq r \leq r_{\rm max} & \mbox{(region $N$)},
\end{array} \right.
\label{eq.3.2.1.2}
\end{equation}
%
%
where $\alpha_{j}$ and $\beta_{j}$ are unknown amplitudes, $A_{R}$ is unknown amplitude of the full reflection,
and $k_{j} = \frac{1}{\hbar}\sqrt{2m(\tilde{E}-V_{j})}$ are complex wave numbers.
We fix a normalization so that the modulus of amplitude of the starting wave $e^{-ik_{N}r}$ equals unity.
We shall search a solution of this problem by the multiple internal reflections approach.

The coefficients $T_{1}^{\pm}$ \ldots $T_{N-1}^{\pm}$ and $R_{1}^{\pm}$ \ldots $R_{N-1}^{\pm}$
can be found from the recurrence relations shown in previous task~\cite{Maydanyuk.2011.JMP}:
\begin{equation}
\begin{array}{ll}
\vspace{2mm}
   T_{j}^{+} = \displaystyle\frac{2k_{j}}{k_{j}+k_{j+1}} \,e^{i(k_{j}-k_{j+1}) r_{j}}, &
   T_{j}^{-} = \displaystyle\frac{2k_{j+1}}{k_{j}+k_{j+1}} \,e^{i(k_{j}-k_{j+1}) r_{j}}, \\
   R_{j}^{+} = \displaystyle\frac{k_{j}-k_{j+1}}{k_{j}+k_{j+1}} \,e^{2ik_{j}r_{j}}, &
   R_{j}^{-} = \displaystyle\frac{k_{j+1}-k_{j}}{k_{j}+k_{j+1}} \,e^{-2ik_{j+1}r_{j}}.
\end{array}
\label{eq.3.2.1.3}
\end{equation}

Each step in such a consideration of packet propagation is similar to one of the first independent $2N-1$ steps.
From the analysis of these steps, we find recurrent relations for the calculation of the unknown amplitudes
$A_{T}^{(n)}$, $A_{R}^{(n)}$, $\alpha_{j}^{(n)}$ and $\beta_{j}^{(n)}$ for an arbitrary step with number $n$
(here, index $j$ corresponds to the number of region $V_{j}$, and the logic of the definition of these amplitudes
can be found in Appendix~A in~\cite{Maydanyuk.2011.JMP}). In the summation of these relations from each step,
we impose the continuity condition for the full (summarized) wave function and its derivative relative to the
corresponding boundary.

Now, let us find a wave propagating to the left in region with number $j-1$ which is formed after transmission through the boundary at point $r_{j-1}$ of all possible incident waves,
produced  in result of all possible reflections and transmissions of any waves in the right part of potential from this boundary.
Amplitude of this wave can be determined as a summation of amplitudes of all waves incident on boundary at point $r_{j-1}$ multiplied on factor $T_{j-1}^{-}$.
It needs to take into account that any wave incident on boundary at $r_{j-1}$ can be reflected from this boundary then can be reflected from the boundary at $r_{j}$ and is incident on boundary at $r_{j-1}$ once again.
We write
\begin{equation}
\begin{array}{lcl}
  \tilde{T}_{j-1}^{-} & = &
    \tilde{T}_{j}^{-} T_{j-1}^{-}
    \Bigl(1 + \sum\limits_{m=1}^{+\infty} (R_{j-1}^{-} \tilde{R}_{j}^{+})^{m} \Bigr) =
    \displaystyle\frac{\tilde{T}_{j}^{-} T_{j-1}^{-}} {1 - R_{j-1}^{-} \tilde{R}_{j}^{+}}.
\end{array}
\label{eq.3.2.1.4}
\end{equation}
Here, we use a summarized reflection amplitude $\tilde{R}_{j}^{+}$
(which should take into account possibility of waves to transfer through boundary at $r_{j}$, propagate to the right,
then after any reflections and transmissions to return back to region with number $j$)
which can be found as
\begin{equation}
\begin{array}{l}
   \vspace{1mm}
   \tilde{R}_{j-1}^{+} =
     R_{j-1}^{+} + T_{j-1}^{+} \tilde{R}_{j}^{+} T_{j-1}^{-}
     \Bigl(1 + \sum\limits_{m=1}^{+\infty} (\tilde{R}_{j}^{+}R_{j-1}^{-})^{m} \Bigr) =
     R_{j-1}^{+} +
     \displaystyle\frac{T_{j-1}^{+} \tilde{R}_{j}^{+} T_{j-1}^{-}} {1 - \tilde{R}_{j}^{+} R_{j-1}^{-}}.
\end{array}
\label{eq.3.2.1.5}
\end{equation}
By such a way, we have obtained recurrent relations which connect all amplitudes.
We choose the following values
\begin{equation}
\begin{array}{cccc}
  \tilde{R}_{N-1}^{+} = R_{N-1}^{+}, & \quad
  \tilde{T}_{N-1}^{-} = T_{N-1}^{-},
\end{array}
\label{eq.3.2.1.6}
\end{equation}
as starting point and consequently calculate all amplitudes
$\tilde{R}_{N-2}^{+}$ \ldots $\tilde{R}_{N_{\rm cap}}^{+}$,
and $\tilde{T}_{N-2}^{-}$ \ldots $\tilde{T}_{N_{\rm cap}}^{-}$.
We define summarized amplitude $A_{T}$ of transition through the barrier via all waves transmitted through the potential region with the barrier from $r_{\rm cap}$ to $r_{N-1}$,
and so we have
\begin{equation}
\begin{array}{lll}
  A_{T, {\rm bar}} = \tilde{T}_{N_{\rm cap}}^{-}. &
\end{array}
\label{eq.3.2.1.7}
\end{equation}
%
In order to find summation of all waves reflected from the boundary at point $r_{j+1}$ and propagating to the right,
we calculate a summarized amplitude of reflection as
\begin{equation}
\begin{array}{lcl}
  \vspace{1mm}
  \tilde{R}_{j+1}^{-} & = &
    R_{j+1}^{-} + T_{j+1}^{-} \tilde{R}_{j}^{-} T_{j+1}^{+}
    \Bigl(1 + \sum\limits_{m=1}^{+\infty} (R_{j+1}^{+} \tilde{R}_{j}^{-})^{m} \Bigr) =
    R_{j+1}^{-} +
    \displaystyle\frac{T_{j+1}^{-} \tilde{R}_{j}^{-} T_{j+1}^{+}} {1 - R_{j+1}^{+} \tilde{R}_{j}^{-}}.
\end{array}
\label{eq.3.2.1.8}
\end{equation}
On such a basis, we define amplitude of reflection from the potential region with the barrier from $r_{\rm cap}$ to $r_{N-1}$ as
\begin{equation}
\begin{array}{lll}
  A_{R, {\rm bar}} = \tilde{R}_{N-1}^{-}, &
  {\rm where}\;
  \tilde{R}_{N_{\rm cap}}^{-} = R_{N_{\rm cap}}^{-}.
\end{array}
\label{eq.3.2.1.9}
\end{equation}

\subsubsection{Amplitudes of reflection from barrier inside tunneling region and external part of barrier
\label{sec.3.2.2}}

We can also find a summarizing amplitude $A_{R, {\rm ext}}$ of all waves reflected from the external barrier region
(from the external turning point $r_{\rm tp,ext}$ to $r_{N-1}$)
and propagated outside (which can characterize a potential scattering) as
\begin{equation}
\begin{array}{lll}
  A_{R, {\rm ext}} = \tilde{R}_{N-1}^{-}, &
  {\rm where}\;
  \tilde{R}_{N_{\rm tp,ext}}^{-} = R_{N_{\rm tp,ext}}^{-}
\end{array}
\label{eq.3.2.2.1}
\end{equation}
and a summarizing amplitude $A_{R, {\rm tun}}$ of all waves which are reflected just inside the potential region from $r_{\rm cap}$ to the external turning point $r_{\rm tp,ext}$
[i.e. they are propagated through the external barrier region (without any reflection), tunnels under the barrier,
may propagate up to the boundary at point $r_{\rm cap}$
and then are reflected back from this boundary]
as
\begin{equation}
\begin{array}{lll}
  A_{R, {\rm tun}} = A_{R, {\rm bar}} - A_{R, {\rm ps}}.
\end{array}
\label{eq.3.2.2.2}
\end{equation}
%
We estimate amplitude of oscillations $A_{\rm osc}$ in region of capture with number $N_{\rm cap}$ as
\begin{equation}
\begin{array}{lllll}
  A_{\rm osc} (N_{\rm cap}) = \displaystyle\frac{1}{1 - \tilde{R}_{N_{\rm cap}-1}^{-} \tilde{R}_{N_{\rm cap}}^{+}}.
\end{array}
\label{eq.3.2.2.3}
\end{equation}
\subsubsection{Coefficients of penetrability and reflection concerning to barrier, test
\label{sec.3.2.3}}

%
We define coefficients of penetrability $T_{\rm bar}$ and reflection $R_{\rm bar}$ concerning to the whole barrier
(i.e. the potential region from $r_{\rm cap}$ to $r_{N-1}$),
and add also definitions for
coefficient $R_{\rm ext}$ of reflection from the external part of barrier
(i.e. region from $r_{\rm tp, ext}$ to $r_{N-1}$),
coefficient $R_{\rm tun}$ of reflection from the barrier region (i.e. region from $r_{\rm cap}$ to $r_{\rm tp, ext}$)
as
\begin{equation}
\begin{array}{ccccc}
  T_{\rm bar} \equiv \displaystyle\frac{k_{\rm cap}}{k_{N}}\; \bigl|A_{T, {\rm bar}}\bigr|^{2}, &
  R_{\rm bar} \equiv \bigl|A_{R, {\rm bar}}\bigr|^{2}, &
  R_{\rm ext} \equiv \bigl|A_{R, {\rm ext}}\bigr|^{2}, &
  R_{\rm tun} \equiv \bigl|A_{R, {\rm tun}}\bigr|^{2}.
\end{array}
\label{eq.3.2.3.1}
\end{equation}
%
We check the property
\begin{equation}
  T_{\rm bar} + R_{\rm bar} = 1,
\label{eq.3.2.3.2}
\end{equation}
%
%
which used as text to indicate whether the MIR method gives the proper solution for the wave function.


\subsubsection{Summations of amplitudes
\label{sec.3.2.4}}

%
Let us calculate sums of the amplitudes $\alpha_{j}^{(i)}$ and $\beta_{j}^{(i)}$.
To determine them, we generalize the formulas (\ref{eq.3.1.3.1}) for the simplest potential.
For the sums of amplitudes for the region with number $j$, we get
\begin{equation}
\begin{array}{ll}
  \beta_{j} \equiv
  \displaystyle\sum\limits_{i=1} \beta_{j}^{(i)} =
  \tilde{T}_{j}^{-}\, \Bigl( 1 + \displaystyle\sum\limits_{i=1} (\tilde{R}_{j-1} \tilde{R}_{j}^{+})^{i} \Bigr) =
  \displaystyle\frac{\tilde{T}_{j}^{-}}{1 - \tilde{R}_{j-1} \tilde{R}_{j}^{+}}, \\

  \alpha_{j} \equiv
  \displaystyle\sum\limits_{i=1} \alpha_{j}^{(i)} =
  \tilde{R}_{j-1}\, \displaystyle\sum\limits_{i=1} \beta_{j}^{(i)} =
  \displaystyle\frac{\tilde{R}_{j-1} \tilde{T}_{j}^{-}}{1 - \tilde{R}_{j-1} \tilde{R}_{1}^{+}}.
\end{array}
\label{eq.3.2.4.1}
\end{equation}

\subsubsection{Probability of existence of the compound nucleus
\label{sec.3.2.5}}

%
To determine the probability of the existence of a compound nucleus, we will look for the integral from the square of the wave function over the inner region to the barrier.
In this paper we define this region via region between two internal turning points $r_{\rm int,1}$ and $r_{\rm int,2}$.
In the region of above-barrier energies, we have:
\begin{equation}
\begin{array}{lll}
  \displaystyle\int\limits_{r_{\rm int,1}}^{r_{\rm int,2}} |\chi(r)|^{2} dr =

  \displaystyle\sum\limits_{j=1}^{n_{\rm int}}
  \displaystyle\int\limits_{r_{j-1}}^{r_{j}}
    \Bigl|
      \displaystyle\sum\limits_{i=1} \alpha_{j}^{(i)} e^{ik_{j}r} +
      \displaystyle\sum\limits_{i=1} \beta_{j}^{(i)} e^{-ik_{j}r} \Bigr|^{2} dr = 

%
%

  \displaystyle\sum\limits_{j=1}^{n_{\rm int}}
  \Bigl\{
    \bigl( |\alpha_{j}|^{2} + |\beta_{j}|^{2} \bigr)\, \Delta r +
    \displaystyle\frac{|\alpha_{j}\beta_{j}|} {k_{j}}\,
      \sin(\theta_{\alpha_{j}} - \theta_{\beta_{j}} + 2k_{j}r)
    \Bigr|_{r_{j-1}}^{r_{j}}
  \Bigr\},
\end{array}
\label{eq.3.2.5.1}
\end{equation}
where $\theta_{\alpha_{j}}$ and $\theta_{\beta_{j}}$ are phases of the amplitudes $\alpha_{j}$ and $\beta_{j}$, respectively.
If the spatial region includes tunneling, then the formulas above should be rewritten in the complex form
(where $k_{j}$ are complex numbers):
%
%
\begin{equation}
\begin{array}{lll}
  \displaystyle\int\limits_{0}^{r_{\rm int,max}} |\chi(r)|^{2}\; dr =

  \displaystyle\sum\limits_{j=1}^{n_{\rm int,max}}
  \displaystyle\int\limits_{r_{j-1}}^{r_{j}}
    \bigl| \alpha_{j} e^{ik_{j}r} + \beta_{j} e^{-ik_{j}r} \bigr|^{2} \;dr =

%

  \displaystyle\sum\limits_{j=1}^{n_{\rm int,max}}
  \Bigl\{
    \bigl( |\alpha_{j}|^{2} + |\beta_{j}|^{2} \bigr)\, \Delta r +
    \displaystyle\frac{\alpha_{j}\beta_{j}^{*}} {2ik_{j}}\,  e^{2ik_{j}r}
      \Bigr|_{r_{j-1}}^{r_{j}} -
    \displaystyle\frac{\alpha_{j}^{*}\beta_{j}} {2ik_{j}}\,  e^{-2ik_{j}r}
      \Bigr|_{r_{j-1}}^{r_{j}}
  \Bigr\}.
\end{array}
\label{eq.3.2.5.2}
\end{equation}
%
%
In this paper we define the probability of the existence of the compound nucleus through the integral (\ref{eq.3.2.5.2}) over the spatial region between two internal turning points
(where the larger internal turning point is determined as the turning point of the barrier for sub-barrier energies, or as
coordinate of the barrier maximum for above-barrier energies):
\begin{equation}
  P_{\rm cn} \equiv
  \displaystyle\int\limits_{r_{\rm int,1}}^{r_{\rm int,2}} |\chi(r)|^{2}\; dr =
  \displaystyle\sum\limits_{j=1}^{n_{\rm int}}
  \Bigl\{
    \bigl( |\alpha_{j}|^{2} + |\beta_{j}|^{2} \bigr)\, \Delta r +
    \displaystyle\frac{\alpha_{j}\beta_{j}^{*}} {2ik_{j}}\,  e^{2ik_{j}r}
      \Bigr|_{r_{j-1}}^{r_{j}} -
    \displaystyle\frac{\alpha_{j}^{*}\beta_{j}} {2ik_{j}}\,  e^{-2ik_{j}r}
      \Bigr|_{r_{j-1}}^{r_{j}}
  \Bigr\}.
\label{eq.3.2.5.3}
\end{equation}
%
%
Comparing this formula with (\ref{eq.3.1.5.4}), we see that it is much more complicated.
It is not possible to represent it exactly as a simple product of three coefficients of permeability, oscillations and localization.

\subsubsection{Cross-section of fusion and coefficients of fusion
\label{sec.3.2.6}}

%
In description of capture of nuclei by nuclei, there is a standard definition of the fusion cross section $\sigma$ based on the barrier penetrability $T_{\rm bar, l}$ and fusion probabilities $P_{l}$
(which takes place as soon as nucleus incident on the barrier, has tunneled through this barrier)
(for example, see Ref.~\cite{Eberhard.1979.PRL}):
\begin{equation}
\begin{array}{lll}
  \sigma_{\rm fus} (E) = \displaystyle\sum\limits_{l=0}^{+\infty} \sigma_{l}(E), &
  \sigma_{l}(E) = \displaystyle\frac{\pi\hbar^{2}}{2mE}\, (2l+1)\, T_{{\rm bar,} l}(E)\, P_{l},
\end{array}
\label{eq.3.2.6.1}
\end{equation}
%
%
where $\sigma_{l}$ is the partial cross-section of fusion at $l$,
$E$ is the energy of the relative motion of the incident nucleus relative to another nucleus.
To study the compound nucleus, we introduce a new definition of the partial fusion cross section in terms of the probability of the existence of a compound nucleus (\ref{eq.3.2.5.3}) as
\begin{equation}
  \sigma_{l} = \displaystyle\frac{\pi\hbar^{2}}{2mE}\, (2l+1)\, f_{l}(E)\, P_{\rm cn} (E),
\label{eq.3.2.6.2}
\end{equation}
%
%
where $f_{l}(E)$ is an additional factor that is needed to connect probability $P_{\rm cn} (E)$, penetrability $T_{{\rm bar,} l}(E)$ and the old factor of fusion $P_{l}$.
To find the explicit form of this coefficient, we consider a case of complete fusion, described by the old formula.
A similar result should give a coefficient at coefficients of fusion equal to one.
We obtain:
\begin{equation}
  f(E) = \displaystyle\frac{k_{\rm cap}}{k_{N}\, |r_{\rm cap} - r_{\rm tp,in, 1}|}.
\label{eq.3.2.6.3}
\end{equation}
%
%
Now, to study formation of the compound nucleus with slow fusion (i.e., without instantaneous fusion),
we vary these fusion coefficients in the spatial region between the capture point $r_{\rm cap}$ and the internal second turning point $r_{\rm int, 2}$.

\section{Analysis
\label{sec.analysis}}

\subsection{Distance between nuclei in stellar matter
\label{sec.analysis}}

%
Let us determine at what distance the nuclei are in the stellar matter, taking the density of the matter into account.
We denote the distance between two close nuclei fixed in the lattice as $2\, R_{0}$, and put a ``scattering'' nucleus between these nuclei
(along to Ref.~\cite{ShapiroTeukolsky.2004.book}, p.~90, Fig.~3.5).
The density $\rho_{0}$ in the sphere surrounding one nucleus of the lattice can be found as
the ratio of mass $m_{A}$ of the nucleus to the volume $V_{A}$ inside this sphere as
\begin{equation}
\begin{array}{llllll}
    \rho_{0} =
    \displaystyle\frac{m_{A}}{V_{A}} =
    \displaystyle\frac{A\, m_{u}}{4/3\, \pi\, R_{0}^{3}}

\end{array}
\label{eq.analysis.2.2}
\end{equation}
or
%
\begin{equation}
\begin{array}{llllll}
  R_{0} = \Bigl( \displaystyle\frac{A\, m_{u}}{4/3\, \pi\, \rho_{0}} \Bigr)^{1/3},
\end{array}
\label{eq.analysis.2.3}
\end{equation}
%
%
where $m_{u}$ is mass of nucleon,
$A$ is mass number of nucleus.
For pycnonuclear reaction $\isotope[12]{C} + \isotope[12]{C} = \isotope[24]{Mg}$ we can choose estimation of the density,
according to Ref.~\cite{ShapiroTeukolsky.2004.book} 
($A=12$)
\begin{equation}
\begin{array}{llllll}
  \rho_{0} = 6 \cdot 10^{9}\, \displaystyle\frac{\mbox{g}} {\mbox{\rm cm}^{3}}.
\end{array}
\label{eq.analysis.2.4}
\end{equation}
%
From Eq.~(\ref{eq.analysis.2.3}) we find
\begin{equation}
\begin{array}{llllll}
  R_{0} = 92.5\; \mbox{\rm fm}
\end{array}
\label{eq.analysis.2.5}
\end{equation}
%
and concentration of nuclei $n_{A}$ is calculated as
\begin{equation}
\begin{array}{llllll}
  \Bigl( \rho_{0} = A\, m_{u} \cdot n_{A} \Bigr) & \to &
  \Bigl( n_{A} = \displaystyle\frac{\rho_{0}}{A\, m_{u}} \Bigr).
\end{array}
\label{eq.analysis.2.6}
\end{equation}
%
Calculations give the following result:
\begin{equation}
\begin{array}{llllll}
  n_{A} = 
  2.411351533\; \mbox{\rm MeV}^{3} =
  3.\: 014\: 189\: 41 \cdot 10^{-7}\; \mbox{\rm fm}^{-3}.
\end{array}
\label{eq.analysis.2.7}
\end{equation}

\subsection{Potential of interaction between nuclei
\label{sec.analysis.3}}

\begin{figure}[htbp]
\centerline{\includegraphics[width=88mm]{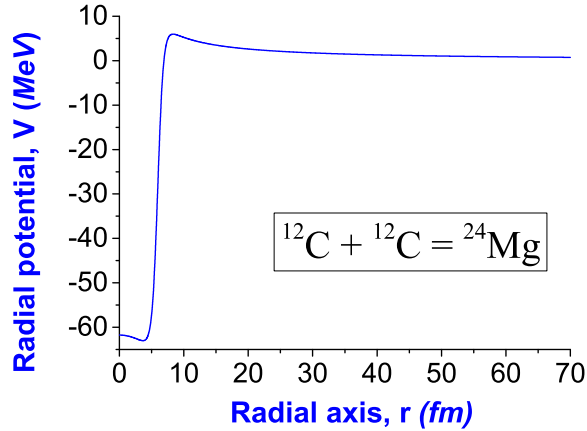}}
\caption{\small (Color online)
Potential of interaction between two nuclei \isotope[12]{C}
[potential and parameters are defined in Eq.~(\ref{eq.analysis.3.1})--(\ref{eq.analysis.3.3})].
%
\label{fig.3.1}}
\end{figure}

We define the potential of interactions between nuclei \isotope[12]{C} as
\begin{equation}
  V (r) = v_{c} (r) + v_{N} (r) + v_{l=0} (r),
\label{eq.analysis.3.1}
\end{equation}
where $v_{c} (r)$, $v_{N} (r)$ and $v_{l} (r)$ are Coulomb, nuclear and centrifugal components
have the form
%
\begin{equation}
\begin{array}{lll}
  \vspace{1mm}
  v_{N} (r) = - \displaystyle\frac{V_{R}} {1 + \exp{\displaystyle\frac{r-R_{R}} {a_{R}}}},
  \hspace{2mm}
  v_{l} (r) = \displaystyle\frac{l\,(l+1)} {2mr^{2}}, \\


  v_{c} (r) =
  \left\{
  \begin{array}{ll}
    \displaystyle\frac{Z e^{2}} {r}, &
      \mbox{at  } r \ge R_{c}, \\
    \displaystyle\frac{Z e^{2}} {2 R_{c}}\;
      \biggl\{ 3 -  \displaystyle\frac{r^{2}}{R_{c}^{2}} \biggr\}, &
      \mbox{at  } r < R_{c}.
  \end{array}
  \right.
\end{array}
\label{eq.analysis.3.2}
\end{equation}
Here, $V_{R}$ is strength of nuclear components defined in MeV as
%
\begin{equation}
\begin{array}{ll}
  V_{R} = -75.0\; \mbox{\rm MeV},
\end{array}
\label{eq.analysis.3.3}
\end{equation}
%
$R_{c}$ and $R_{R}$ are Coulomb and nuclear radiuses of di-nuclear system, $a_{R}$ is diffusion parameter.
In calculations we use%
\footnote{Such a potential with parameters was used by us previously for tests in calculations of interaction between nuclear fragments of middle and heavy masses
in study of bremsstrahlung emission in fission~\cite{Maydanyuk.2010.PRC,Maydanyuk.2010.IJMPE,Maydanyuk.2011.JPCS}. 
So, it is easy way just to continue formalism and codes in new problem.
One can find in literature more accurate formulations of the potential of interactions of these nuclei.
For example, in Ref.~\cite{Afanasjev.2007.PRC} $S$ factors are computed on the basis of the Sao Paulo potential for a
broad range of fusion reactions involving different isotopes
(see also more precise method taking into account more physical properties of nuclei in Ref.~\cite{Denisov.2015.PRC}).
However, such formulations include many details related with determination of the potential,
which do not take into account analysis of quantum fluxes studied in this paper.
So, it is more effective on current stage to describe our quantum method without technical complications of the potential and related computations.}
\begin{equation}
\begin{array}{llll}
  R_{R} = r_{R}\, (A_{1}^{1/3} + A_{2}^{1/3}), &
  R_{c} = r_{c}\, (A_{1}^{1/3} + A_{2}^{1/3}), &
  a_{R} = 0.44\; {\rm fm}, \\
  r_{R} = 1.30\;{\rm fm}, &
  r_{c} = 1.30\;{\rm fm}.
\end{array}
\label{eq.analysis.3.4}
\end{equation}
%
This potential is shown in Fig.~\ref{fig.3.1}.
%
One can see a shape of internal well of the potential
(in contract to calculations in Ref.~\cite{ShapiroTeukolsky.2004.book}, for example).
We find minimum of the internal well and maximum of the barrier as
%
\begin{equation}
\begin{array}{llllll}
\vspace{1mm}
  r_{\rm min} = 3.643\; \mbox{\rm fm}, & V_{\rm min} = -63.018\; \mbox{MeV}, \\
  r_{\rm max} = 8.338\; \mbox{\rm fm}, & V_{\rm max} = +5.972\; \mbox{MeV}.
\end{array}
\label{eq.analysis.3.5}
\end{equation}

\subsection{Energy of zero-point vibrations in lattice sites and collision of nuclei
\label{sec.analysis.4}}

Energy of zero-point vibrations of nuclei in lattice sites, which is chosen to study collision between the incident nucleus and the nucleus in lattice, is calculated as
\begin{equation}
\begin{array}{llllll}
\vspace{1mm}
  E_{0}  =
  \displaystyle\frac{\hbar w}{2} =
  \displaystyle\frac{\hbar\, Ze}{\sqrt{m\, R_{0}^{3}}}, &
  \Delta E = \displaystyle\frac{2\, Z^{2}e^{2}}{R_{0}}, &
  E_{\rm full} =
  E_{0} + \Delta E.
\end{array}
\label{eq.analysis.4.1}
\end{equation}
%
For $\isotope[12]{C} + \isotope[12]{C} = \isotope[24]{Mg}$ we obtain
\begin{equation}
\begin{array}{llllll}
\vspace{1.5mm}
  E_{0} = 0.021 808 06\; \mbox{\rm MeV}, &
\vspace{1.5mm}
  \Delta E = 0.567 872 37\; \mbox{\rm MeV}, &
  E_{\rm full}^{\rm (zero\, mode)} = 0.589 680 43\; \mbox{\rm MeV}.
\end{array}
\label{eq.analysis.4.2}
\end{equation}
%
For brevity of analysis, we call such a state (and the corresponding energy) as state in zero mode.

\subsection{Penetrability of the barrier, rate of reaction and role of new quantum effects
\label{sec.analysis.5}}

\begin{figure}[htbp]
\centerline{\includegraphics[width=88mm]{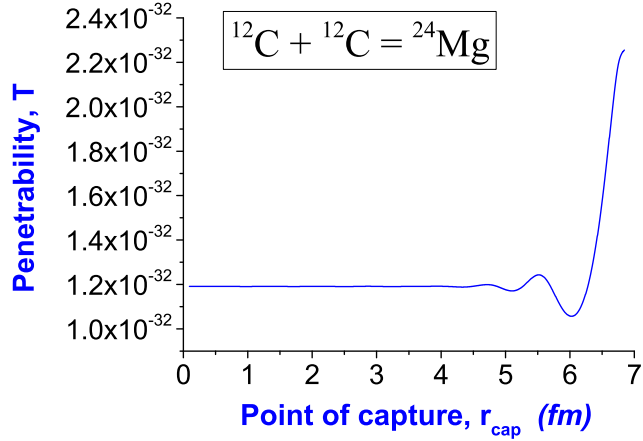}}
\caption{\small (Color online)
Penetrability in dependence on position of the coordinate of maximum fusion in the internal region of the potential
[potential and parameters are defined in Eq.~(\ref{eq.analysis.3.1})--(\ref{eq.analysis.3.3}),
calculation was performed at energy $E=0.58$~MeV,
its excess over the potential is $\Delta E = 0.02$~MeV at point $R_{0}$, see Eq.~(\ref{eq.analysis.4.2})].
\label{fig.3.2}}
\end{figure}

%
Let us calculate the penetrability of such a potential barrier using the method of multiple internal reflections (MIR).
Fig.~\ref{fig.3.2} shows change of the penetrability depending on the displacement of the point where transition to stage with fusion in the internal nuclear reaction can occur.
%
%
The calculation is performed at energy $E=0.58$~MeV (excess of this energy over the potential is $\Delta E = 0.02$~MeV at $r=R_{0}$).
In the figure the extreme right part of the line corresponds to the internal (second) turning point,
i.e. case when the left part of the region of potential (used in calculation of the penetrability) is restricted by the left boundary of the tunneling region.
That corresponds to calculations in the WKB approximation (without taking into account the influence of the external region of potential, which is much smaller), as well as calculations in other approaches.
However, fusion process occurs more probably in the nuclear region after some distance from this turning point.
This consideration describes more properly the internal quantum fluxes in nucleus.
In particular, it improves essentially description of experimental data for related reactions,
where a detailed analysis was performed
(see Refs.~\cite{Maydanyuk.2015.NPA,Maydanyuk_Zhang_Zou.2017.PRC}, for details, demonstrations, references therein).
The difference is
\begin{equation}
\begin{array}{llllll}
\vspace{2.0mm}
  {\rm WKB-approach:} &
  r_{\rm cap} = r_{\rm tp,\,2} = 6.845\, {\rm fm}, &
  T_{\rm bar}^{\rm (WKB)} = 2.255 \cdot 10^{-32}, \\

  {\rm MIR-approach:} &
  r_{\rm cap} = r_{\rm min} = 3.643\, {\rm fm}, &
  T_{\rm bar}^{\rm (MIR)} = 1.191 \cdot 10^{-32}.
\end{array}
\label{eq.analysis.5.1}
\end{equation}
%
%
Unlike previous approaches, now the penetrability depends on the position of fusion process.
One can see that such a shift in the position of the fusion reduces the penetrability by about a factor of two.
An accurate estimation gives:
$T_{\rm bar}^{\rm (WKB)} / T_{\rm bar}^{\rm (MIR)} = 2.255 \cdot 10^{-32} / 1.191 \cdot 10^{-32} = $1.893366.

According to Eq.~(\ref{eq.4.2.4}), we calculate reaction rate as
\begin{equation}
\begin{array}{llllll}
\vspace{2.0mm}
  W_{\rm new} & = &
  \displaystyle\frac{k_{1}\, \hbar}{m} \times \displaystyle\frac{S(E)}{E} \times T_{\rm bar}^{\rm (MIR)} (r_{\rm cap}) =
  \displaystyle\frac{W_{\rm old}}{1.893366}.
\end{array}
\label{eq.analysis.5.2}
\end{equation}
%
%
For calculation one can take estimation of $S_{\rm CC}$ from Ref.~\cite{ShapiroTeukolsky.2004.book} [see p.~95]:
\begin{equation}
\begin{array}{llllll}
  S_{\rm CC} & = & 8.83 \cdot 10^{16}\; \mbox{\rm MeV} \cdot \mbox{\rm barn}.
\end{array}
\label{eq.analysis.5.3}
\end{equation}
%
%
Note that more accurate calculations of the $S$-factors in dense nuclear matter are made in, for example, Ref.~\cite{Yakovlev.2006.PRC}
[see Fig.~1 in this paper;
see also Refs.~\cite{Afanasjev.2012.PRC,Afanasjev.2010.ADNDT,Afanasjev.2007.PRC,Afanasjev.2006.PRC}].
It also provides convenient formulas for self-analysis for other researchers, and we could move forward that way too.
However, to perform the first estimates, $S$-factors in Eq.~(\ref{eq.analysis.5.3}) and such papers are close enough.
Therefore, we will use estimation in Eq.~(\ref{eq.analysis.5.3}) for the $S$ factor in analysis.
%
From Eq.~(\ref{eq.analysis.5.2}) we find
%
%
%
%
%
\begin{equation}
\begin{array}{llllll}
  W_{\rm old} =
    1.943542 \cdot 10^{3}\; \displaystyle\frac{\mbox{\rm MeV}^{-3}}{\mbox{sec}}, &

  W_{\rm new} =
    1.026506 \cdot 10^{3}\; \displaystyle\frac{\mbox{\rm MeV}^{-3}}{\mbox{sec}}.
\end{array}
\label{eq.analysis.5.7}
\end{equation}
%
From Eq.~(\ref{eq.4.3.1}) we find number of reactions 
per second:
\begin{equation}
\begin{array}{llllll}
  P_{\rm old} =
    4.686562 \cdot 10^{3}\; \displaystyle\frac{1}{\mbox{sec}}, &
  P_{\rm new} =
    2.475266 \cdot 10^{3}\; \displaystyle\frac{1}{\mbox{sec}}.
\end{array}
\label{eq.analysis.5.8}
\end{equation}

\subsection{Renormalization of wave function
\label{sec.analysis.6}}

Estimations of rates of the reactions are strongly depended on how we normalize the wave function.
In this topic estimations in unite of volume (usually in cubic centimeters) are used.
So, we renormalize our calculations.
Following to Ref.~\cite{ShapiroTeukolsky.2004.book} [see Eq.~(3.7.22), (3.7.25), p.~92],
square of modulus of the wave function for incident wave on the barrier is normalized on volume of sphere defined by radius $r_{0}$ of above-barrier region as
\begin{equation}
\begin{array}{llllll}
  \bigl| \chi_{\rm inc} \bigr|^{2} =
  \displaystyle\frac{1}{r_{0}^{3}\, \pi^{3/2}}, &
  r_{0} =
  \Bigl( \displaystyle\frac{\hbar}{2\, Ze} \Bigr)^{1/2}\,
  \Bigl( \displaystyle\frac{R_{0}^{3}}{m} \Bigr)^{1/4}.
\end{array}
\label{eq.analysis.6.3}
\end{equation}
%
Calculation gives the following renormalization coefficient
\begin{equation}
\begin{array}{llllll}
  \displaystyle\frac{r_{0}}{R_{0}} = 0.138 569 66, &

  c_{0} =
  \Bigl( \displaystyle\frac{R_{0}}{r_{0}} \Bigr)^{3} \cdot
  \displaystyle\frac{1}{R_{0}^{3}\, \pi^{3/2}} =
  681.741\: 719\; \mbox{MeV}^{3},
\end{array}
\label{eq.analysis.6.5}
\end{equation}
and the following renormalized rates and numbers of reactions (denoted by a tilde above)
\begin{equation}
\begin{array}{llllll}
  \bar{W} = W \cdot c_{0}, &
  \bar{P} = P \cdot c_{0}.
\end{array}
\label{eq.analysis.6.6}
\end{equation}
Finally, we obtain rate of reaction per second and number of reactions per one cubic centimeter per second as
\begin{equation}
\begin{array}{llllll}
\vspace{2.0mm}
  \bar{W}_{\rm old} =
    1 324 993.664\; \displaystyle\frac{1}{\mbox{sec}}, &

  \bar{W}_{\rm new} =
    699 811.965\; \displaystyle\frac{1}{\mbox{sec}}, \\

  \bar{P}_{\rm old} =
    2.556 019 867\; \displaystyle\frac{10^{52}}{\mbox{sec} \cdot \mbox{\rm cm}^{3}}, &

  \bar{P}_{\rm new} =
    1.349 993 677\; \displaystyle\frac{10^{52}}{\mbox{sec} \cdot \mbox{\rm cm}^{3}}.
\end{array}
\label{eq.analysis.6.7}
\end{equation}
\subsection{Resonances and formation of compound nuclear system
\label{sec.analysis.9}}

\begin{figure}[htbp]
\centerline{\includegraphics[width=88mm]{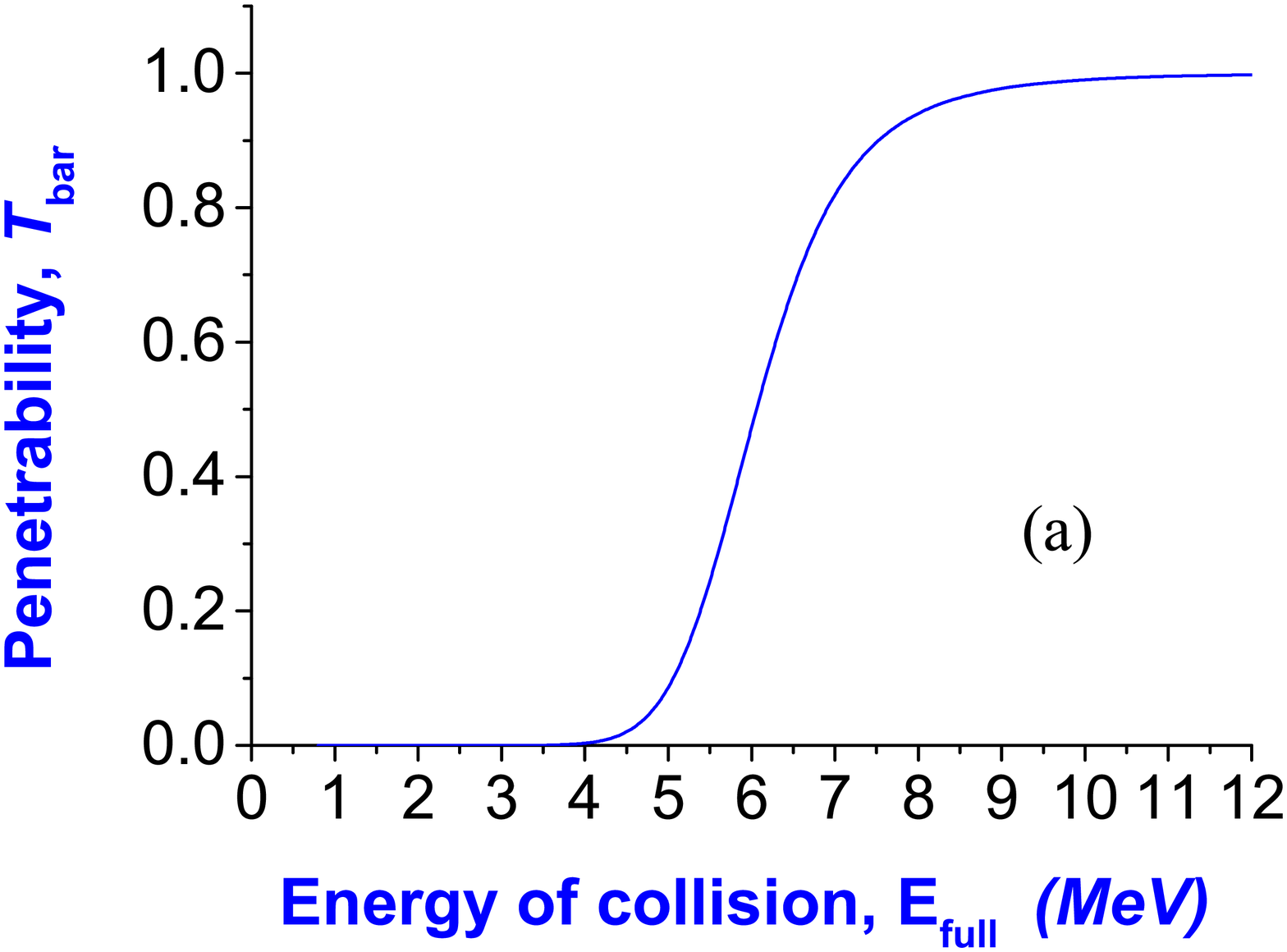}
\hspace{-1mm}\includegraphics[width=88mm]{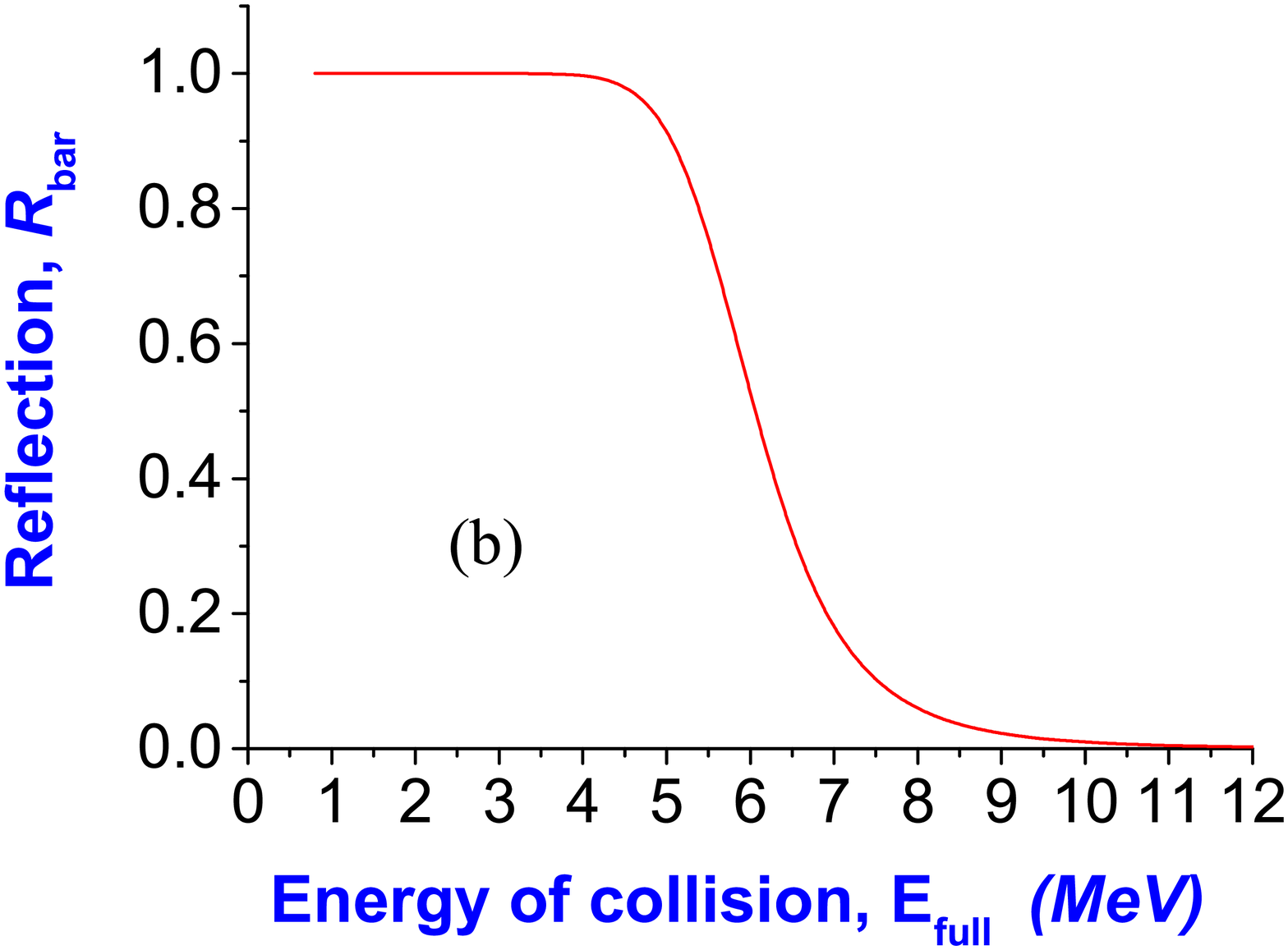}}
\caption{\small (Color online)
%
Coefficients of penetrability (a) and reflection (b) in dependence on the energy of the incident nucleus
for reaction $\isotope[12]{C} + \isotope[12]{C} = \isotope[24]{Mg}$
[coefficients are calculated independently;
potential and parameters are defined in Eq.~(\ref{eq.analysis.3.1})--(\ref{eq.analysis.3.3})].
\label{fig.3.3}}
\end{figure}

We will analyze how the coefficients of penetrability and reflection are changed in dependence on the energy of the incident nucleus.
Calculation of these characteristics is shown in Fig.~\ref{fig.3.3}.
%
%
One can see that both coefficients are changed monotonically in dependence on the energy.
There are no maximums and minimums in such dependencies.
This picture is in full agreement with previous studies on the capture of $\alpha$ particles by
nuclei~\cite{Maydanyuk.2015.NPA,Maydanyuk_Zhang_Zou.2017.PRC}.

\begin{figure}[htbp]
\centerline{\includegraphics[width=88mm]{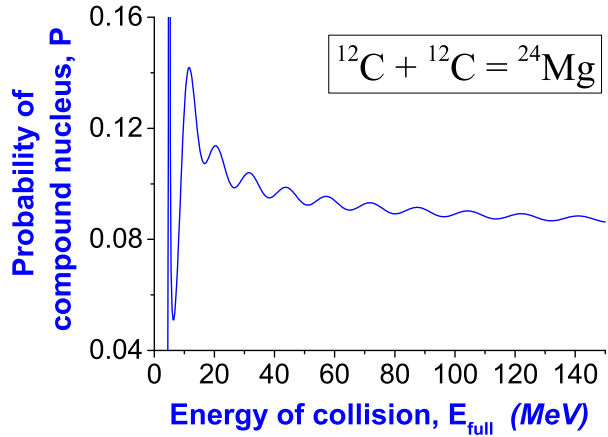}}
\caption{\small (Color online)
%
Dependence of the probability of formation of a compound nucleus $P_{\rm cn}$ on energy of the incident nucleus on the nucleus of lattice
for reaction $\isotope[12]{C} + \isotope[12]{C} = \isotope[24]{Mg}$
[potential and parameters are defined in Eq.~(\ref{eq.analysis.3.1})--(\ref{eq.analysis.3.3})].
Note that the first maximum corresponds to the sub-barrier energy, where the pycnonuclear reaction proceeds through the tunneling stage under the barrier
[see parameters of barrier in Eq.~(\ref{eq.analysis.3.5})].
\label{fig.3.4}}
\end{figure}

Formalism of the method MIR determines also other quantum characteristics.
Probability of formation of a compound nucleus calculated by such a way is shown in Fig.~\ref{fig.3.4}.
%
%
Here, resonance maxima are already clearly observed in the probability curve, which correspond to certain quite definite energies.
This result gives an entirely new picture for processes similar to pycnonuclear reactions.
These maxima are explained by the need to take into account the further propagation of quantum fluxes in the potential region,
in contrast to the existed modern description of pychonuclear reactions, where these fluxes are interrupted and disappeared at internal turning point.
Quantum mechanics requires such a consideration, which strictly indicates on continuity of the wave function in the full region of its definition.
It requires conservation of fluxes in the full region of definition of the wave function
(while calculations of cross-sections and rates of reactions in old approach,
for example, $T^{\rm (WKB)}_{\rm bar}$ by Eq.~(\ref{eq.analysis.5.1}),
$\bar{W}_{\rm old}$, $\bar{P}_{\rm old}$ by Eq.~(\ref{eq.analysis.6.7}),
are based on interruption of the quantum flux at turning point).
The maxima in the probability curve indicate on some new states
in which the compound nucleus is formed and can exist with the maximal probability.
Exactly in such states and at such energies the process of further fusion with formation of the new \isotope[24]{Mg} nucleus is much more probable,
than at energies determined by the theory of pycnonuclear reactions in stars studied else Zel'dovich.

The first energies for such resonant maxima are shown in Tabl.~\ref{table.1}.
\begin{table}
\begin{center}
\begin{tabular}{|c|c|c|c|c|c|c|} \hline
  No. & Energy $E_{\rm full}$, MeV & $P_{cn}$ & $T_{\rm bar}$ & $R_{\rm bar}$ & $R_{\rm pot.\, scat.}$ & $R_{\rm res.\, scat.}$ \\ \hline
  1   &  5.0032 & 0.78050 & 0.08112 & 0.91888 & 0.03581 & 0.61165 \\
  2   & 11.6076 & 0.14195 & 0.99697 & 0.00303 & $6.2445 \cdot 10^{-5}$ & 0.00389 \\
  3   & 20.3134 & 0.11369 & 0.99996 & $3.6922 \cdot 10^{-5}$ & $2.0760 \cdot 10^{-6}$ & $2.2511 \cdot 10^{-5}$ \\
  4   & 31.4208 & 0.10401 & 0.99999 & $8.7746 \cdot 10^{-7}$ & $2.3113 \cdot 10^{-7}$ & $1.9218 \cdot 10^{-6}$ \\
  5   & 43.7290 & 0.09875 & 0.99999 & $4.8371 \cdot 10^{-8}$ & $4.9908 \cdot 10^{-8}$ & $1.7137 \cdot 10^{-8}$ \\
  6   & 57.2380 & 0.09547 & 0.99999 & $3.4622 \cdot 10^{-9}$ & $1.5731 \cdot 10^{-8}$ & $3.2376 \cdot 10^{-8}$ \\
  7   & 71.6477	& 0.09319 & 0.99999 & $2.2762 \cdot 10^{-10}$ & $5.7043 \cdot 10^{-9}$ & $3.7514 \cdot 10^{-9}$ \\
  8   & 87.2581 & 0.09152 & 0.99999 & $1.9270 \cdot 10^{-10}$ & $2.2821 \cdot 10^{-9}$ & $1.3541 \cdot 10^{-9}$ \\
  \hline
\end{tabular}
\end{center}
\caption{%
%
The first resonant states at energies corresponding to the maxima of the probability of formation of a compound nucleus
for reaction $\isotope[12]{C} + \isotope[12]{C} = \isotope[24]{Mg}$, which is shown in Fig.~\ref{fig.3.4}.
Here, $P_{\rm cn}$ is probability of compound nucleus formation defined in Eq.~(\ref{eq.3.2.5.3}),
$T_{\rm bar}$ and $R_{\rm bar}$ are coefficients of penetrability and reflection defined in Eq.~(\ref{eq.3.2.3.1}),
$R_{\rm pot.\, scat.}$ is potential scattering concerning to potential defined by the first term $\alpha_{j}^{(1)}$ in summation in $R_{\rm bar}$,
$R_{\rm res.\, scat.}$ is resonant scattering concerning to potential defined by all terms $\alpha_{j}^{(n)}$ with exception of $\alpha_{j}^{(1)}$ in summation in $R_{\rm bar}$
(index $j$ indicates the external turning point of barrier for under-barrier energies or the maximum of barrier for above-barrier energies)
[for each calculation, we check test of $|T_{\rm bar} + R_{\rm bar}| = 1$ and obtain accuracy of $10^{-14}$ of its confirmation].}
\label{table.1}
\end{table}
%
Let us write down how penetrabilities are different at these two states ($r_{\rm cap} = r_{\rm min} = 3.643\, {\rm fm}$):
\begin{equation}
\begin{array}{llllll}
\vspace{2.0mm}
  E_{\rm full}^{\rm (zero\, mode)} = 0.58\; \mbox{\rm MeV}: &
  T_{\rm bar}^{\rm (zero\, mode)} = 1.191 \cdot 10^{-32}, \\

\vspace{2.0mm}
  E_{\rm full}^{\rm (quasi-bound)} = 5.00\; \mbox{\rm MeV}: &
  T_{\rm bar}^{\rm (quasi-bound)} = 0.08112.
\end{array}
\label{eq.analysis.9.1}
\end{equation}
%
So, we find ratio
\begin{equation}
\begin{array}{llllll}
  \displaystyle\frac
    {T_{\rm bar}^{\rm (quasi-bound)} (E_{\rm full} = 5.00\; \mbox{\rm MeV})}
    {T_{\rm bar}^{\rm (zero\, mode)} (E_{\rm full} = 0.58\; \mbox{\rm MeV})} =
  \displaystyle\frac{0.08112}{1.191 \cdot 10^{-32}} =
  6.811\: 083\: 123 \cdot 10^{30}.
\end{array}
\label{eq.analysis.9.2}
\end{equation}
%
%
According to Eq.~(\ref{eq.4.3.1}), rate and number of pycnonuclear reactions (per unit time per unit volume),
passing through the most probable stage of formation of the compound nucleus will be as many times larger
than these characteristics for zero-point vibrations of nuclei in lattice sites.
So, essentially more reactions proceed through such stable formations with the nuclei in the lattice.
Note that such an effect has not been studied and found yet.

\section{Conclusions and perspectives
\label{sec.conclusion}}

%
A new quantum method for description of pycnonuclear reactions in the compact stars is developed in this paper.
The method is based on the formalism of the method of multiple internal reflections, previously developed for study of quantum processes with high precision in the problems of nuclear decays~\cite{Maydanyuk.2011.JPS,Maydanyuk.2000.UPJ,Maydanyuk.2002.JPS,Maydanyuk.2006.FPL},
nuclear captures by nuclei~\cite{Maydanyuk.2015.NPA,Maydanyuk_Zhang_Zou.2017.PRC},
as well as problems of quantum cosmology, where the idea of tunneling is investigated \cite{Maydanyuk.2011.EPJP}.
Main conclusions of the analysis on the example of the reaction $\isotope[12]{C} + \isotope[12]{C} = \isotope[24]{Mg}$ are the following.

\begin{itemize}
\item
Rates of pycnonuclear reactions are changed essentially, after taking into account nuclear part of the potential of interaction.

\item
%
A fully quantum analysis reduces rate of reactions and number of reactions by 1.8 times [see Eq.~(\ref{eq.analysis.6.7})].
This is explained by that the most probable fusion of the nuclei does not occur immediately after leaving nuclear fragment from the tunnel region (the second internal turning point),
but at propagation at the middle of the internal potential well.

\item
%
%
A fully quantum consideration of the pycnuclear reaction requires a complete analysis of quantum fluxes in the internal region of potential (i.e. in the nuclear system)
(quantum mechanics defines wave function inside the full region of its definition, which must be continuous).
This leads to the appearance of new states in which formation of a compound nuclear system is the most probable (see Fig.~\ref{fig.3.4}, Tabl.~\ref{table.1}).
We call such states as \emph{quasi-bound states in pycnonuclear reactions}
(following to logic and formalism in Refs.~\cite{Maydanyuk.2015.NPA,Maydanyuk_Zhang_Zou.2017.PRC}).
As shown in Fig.~\ref{fig.3.4}, at energies corresponding to such states, such a reaction is essentially more probable, than at energies predicted by
Zel'dovich~\cite{Zeldovich.1965.AstrJ} and developed by followers of this idea.
Therefore, there is a sense to tell about reaction rates for such quasi-bound states, rather than for states of zero-point vibrations in lattice sites.
This leads to the essential changes in estimation of the rates of pycnonuclear reactions in stars [see Eq.~(\ref{eq.analysis.9.2})].
These states of nuclei \isotope[12]{C} are not included to systematic description~\cite{Kelley.2017.NPA}.
There is interesting perspective to test the method presented in this paper on the basis of new experimental measurements in Ref.~\cite{Fang.2017.PRC}.

\end{itemize}

\section*{Acknowledgements
\label{sec.acknowledgements}}

S.~P.~Maydanyuk thanks
the Wigner Research Center for Physics in Budapest
for warm hospitality and support.
Authors are highly appreciated to
Prof. A.~G.~Magner for useful discussions concerning to different aspects of nuclear matter in conditions of compact stars and in Earth,
Prof.~V.~S.~Vasilevsky for useful discussions concerning the method of multiple internal reflections and useful recommendation to include the method of complex scaling to formalism and analysis.

%


\appendix
\section{Pycnonuclear reactions
\label{sec.4}}

\subsection{Rates of reactions and fusion probabilities
\label{sec.4.1}}

%
In pycnonuclear reactions, some additional characteristics are determined.
We follow formalism in Ref.~\cite{ShapiroTeukolsky.2004.book} [see p.~86--94 in that book].
The reaction rate (probability per second) in scattering of one nucleus with charge $Z_{1}$ on another nucleus with charge $Z_{2}$ is
\begin{equation}
\begin{array}{llllll}
\vspace{1.5mm}
  W & = &
  (\mbox{\rm transmitted flux at }\, R_{n}) \times 4\, \pi R_{n}^{2} \times P_{n} = \\

  & = &
  (\mbox{\rm incident flux })\, \times T_{\rm full} \times 4\, \pi R_{n}^{2} \times P_{n}.
\end{array}
\label{eq.4.1.1}
\end{equation}
%
Here $P_{n}$ is the probability of nuclear reaction, starting from moment when incident nucleus has already penetrated the barrier into the inner region of the other nucleus;
$T_{\rm full}$ is coefficient of penetrability of the barrier.

The reaction rate is determined using the reaction cross section $\sigma$ as
\begin{equation}
\begin{array}{llllll}
  W = \sigma (E) \times (\mbox{\rm incident flux}).
\end{array}
\label{eq.4.1.2}
\end{equation}
%
Here, $E$ is energy of scattering of nucleus on the other nucleus.
Fhe following formula of cross-section of reaction is used traditionally in this topic
(for example, see Ref.~\cite{ShapiroTeukolsky.2004.book}, Eq.~(3.7.15), p.~90):
\begin{equation}
\begin{array}{llllll}
  \sigma (E) & = & \displaystyle\frac{S(E)}{E} \times T_{\rm full},
\end{array}
\label{eq.4.1.3}
\end{equation}
%
that gives
\begin{equation}
\begin{array}{llllll}
\vspace{2.0mm}
  W & = &
  \displaystyle\frac{4\pi R_{n}^{2}\, P_{n} \, E}{E} \times T_{\rm full} \times (\mbox{\rm incident flux}) =
  \displaystyle\frac{S(E)}{E} \times T_{\rm full} \times (\mbox{\rm incident flux}), \\

  S (E) & = & 4\pi R_{n}^{2}\, P_{n} \, E,
\end{array}
\label{eq.4.1.4}
\end{equation}

%
Now we repeat these calculations, but on the basis of method of multiple internal reflections described above.
Previously cross-section (probability) for formation of a compound nucleus in the scattering of the $\alpha$ particles on nuclei was determined
[it can be associated with the processes of capture of $\alpha$ particles by nuclei]
on the basis of the barrier penetrability $T_{\rm bar, l}$ and probabilities of fusion $P_{l}$ (which takes place as soon as projectile has tunneled through the barrier):
\begin{equation}
\begin{array}{lll}
  \sigma_{\rm fus} (E) = \displaystyle\sum\limits_{l=0}^{+\infty} \sigma_{l}(E), &
  \sigma_{l} (E) = \displaystyle\frac{\pi\hbar^{2}}{2mE}\, (2l+1)\, f_{l}(E)\, P_{\rm cn} (E),
\end{array}
\label{eq.4.1.5}
\end{equation}
%
where [see Eq.~(\ref{eq.3.1.5.4})]
\begin{equation}
\begin{array}{llllll}
  P_{\rm cn} =
    P_{\rm osc}\, T_{\rm bar}\, P_{\rm loc}, & \\

  P_{\rm osc} = |A_{\rm osc}|^{2} =
  \displaystyle\frac{(k + k_{1})^{2}}
    {2k^{2} (1 -\cos (2k_{1}r_{1})) + 2k_{1}^{2}\,(1 + \cos (2k_{1}r_{1})) }, & \\

  T_{\rm bar} \equiv \displaystyle\frac{k_{1}}{k_{2}}\; \bigl| T_{1}^{-} \bigr|^{2}, & \\

  P_{\rm loc} = 2\, \displaystyle\frac{k_{2}}{k_{1}}\; \Bigl( r_{1} - \displaystyle\frac{\sin(2k_{1}r_{1})}{2k_{1}} \Bigr).
\end{array}
\label{eq.4.1.6}
\end{equation}
%
Here, $\sigma_{l}$ is the partial cross-section of capture at $l$,
$T_{\rm bar}$ is the barrier penetrability (concerning to the strictly chosen coordinate of fusion, without oscillations).

%
One can use $f_{l}(E)\, P_{\rm cn}$ instead of $T_{\rm full}$ ($T_{\rm full} \to f_{l}(E) \cdot P_{\rm cn}$).
From Eq.~(\ref{eq.4.1.4}) and (\ref{eq.4.1.5}) we write
(we use $l=0$, following to Ref.~\cite{ShapiroTeukolsky.2004.book})
\begin{equation}
\begin{array}{llllll}
  W =
  \displaystyle\frac{S(E)}{E} \times f_{l}(E)\, P_{\rm cn} \times (\mbox{\rm incident flux}), &

  S (E) = \displaystyle\frac{\pi\hbar^{2}}{2m}\, (2l+1).
\end{array}
\label{eq.4.1.7}
\end{equation}
%
%
%
%
Here, $f_{l}(E)$ is an additional coefficient that is needed to connect probability $P_{\rm cn} (E)$, penetrability $T_{{\rm bar,} l}(E)$ and the old factor $P_{l}$.
To find explicit value of this coefficient, we consider a case of fast complete fusion, described by the old formula.
A similar result gives the coefficient of fusion to be equal to unite.
We get:
\begin{equation}
  f(E) = \displaystyle\frac{k_{\rm cap}}{k_{N}\, |r_{\rm cap} - r_{\rm tp,in, 1}|}.
\label{eq.4.1.8}
\end{equation}
%
%
Let us calculate this coefficient for the potential of a simple form shown in Fig.~\ref{fig.2.1}:
\begin{equation}
  f^{\rm (simple)}(E) = \displaystyle\frac{k_{1}}{k_{2}\, r_{1}},
\label{eq.4.1.9}
\end{equation}
%
where
\begin{equation}
\begin{array}{llllll}
  k_{\rm cap} = k_{1}, & k_{N} = k_{2}, & r_{\rm cap} = r_{1}, & r_{\rm tp,in, 1} = 0.
\end{array}
\label{eq.4.1.10}
\end{equation}

\subsection{Quantum flux and number of pycno-nuclear reactions
\label{sec.4.2}}

To find flux, we use its definition in quantum mechanics.
According to Ref.~\cite{Landau.v3.1989} (see Eq.~(19.4), p.~80 in this book), we define the flux as
\begin{equation}
  j\,(r) = \displaystyle\frac{i\, \hbar}{2m}\, (\varphi \nabla \varphi^{*} - \varphi^{*} \nabla \varphi).
\label{eq.4.2.1}
\end{equation}
%
Then, for the incident flux (i.e. the flux for the incident wave in the form of $\exp(-ik_{1}r)$) we have ($r > r_{1}$)
\begin{equation}
\begin{array}{llllll}
  j_{\rm inc}\,(r) & = &
  \displaystyle\frac{i\, \hbar}{2m}\,
    \Bigl[\exp(-ik_{1}r) \displaystyle\frac{d}{dr} \exp(-ik_{1}r)^{*} - \exp(-ik_{1}r)^{*} \displaystyle\frac{d}{dr} \exp(-ik_{1}r) \Bigr] =



  - \displaystyle\frac{k_{1}\, \hbar}{m}.
\end{array}
\label{eq.4.2.2}
\end{equation}
%
%
We write formula for the reaction rate as
\begin{equation}
\begin{array}{llllll}
  W & = &
  \displaystyle\frac{k_{1}\, \hbar}{m} \times \displaystyle\frac{\pi\hbar^{2}}{2mE}\, (2l+1)\, f_{l}(E)\, P_{\rm cn} (E) \Bigl|_{l=0} =
  \displaystyle\frac{k_{1}\, \hbar}{m} \times \displaystyle\frac{S(E)}{E} \times f_{l}(E)\, P_{\rm cn}.

\end{array}
\label{eq.4.2.3}
\end{equation}
%
Here, there are formulas in the old and new definitions:
\begin{equation}
\begin{array}{llllll}
\vspace{2.0mm}
  W_{\rm old} & = &
  \displaystyle\frac{k_{1}\, \hbar}{m} \times \displaystyle\frac{S(E)}{E} \times T_{\rm bar}, \\

\vspace{2.0mm}
  W_{\rm new} & = &
  \displaystyle\frac{k_{1}\, \hbar}{m} \times \displaystyle\frac{S(E)}{E} \times f_{l}(E)\, P_{\rm cn}, &

  P_{\rm cn}^{\rm (fast\, fusion)} = T_{\rm bar} \times P_{\rm osc}\, P_{\rm loc}.
\end{array}
\label{eq.4.2.4}
\end{equation}
%
%
We determine number of reactions per cubic centimeter per second, following formalism in
Ref.~\cite{ShapiroTeukolsky.2004.book} [see~p.~94, Eq.~(3.7.38) in that book]:
\begin{equation}
  P_{0} = n_{A}\, W,
\label{eq.4.3.1}
\end{equation}
%
where $n_{A}$ is concentration of nuclei (in crystal lattice), it is estimated under assumption that one nucleus is in the sphere with radius $R_{0}/2$.

For analysis we will choose reaction of $\isotope[12]{C} + \isotope[12]{C} = \isotope[24]{Mg}$~\cite{Gasques.2005.PRC}. 
Different estimations of density were obtained for such a reaction (for example, see Ref.~\cite{ShapiroTeukolsky.2004.book}, p.~83).
Hamada and Salpeter estimated that
\isotope[12]{C} would be converted to \isotope[24]{Mg} via pycnonuclear reactions above a density of $6 \cdot 10^{9}$ ${\rm g} \cdot {\rm cm}^{-3}$
\cite{Hamada_Salpeter.1961.AstrJ},
that was based on estimations of pycnonuclear reaction rates calculated by Cameron~\cite{Cameron.1959b.AstrJ}. 
Those estimations then were improved (for example, see calculations by Salpeter and Van Horn~\cite{Salpeter_VanHorn.1969.AstrJ}, etc.).
However, the densities quoted here are still quite uncertain.
Besides the difficulty of an accurate calculation, finite temperatures and crystal imperfections can increase the rates significantly
(the critical density for carbon can be about $5 \cdot 10^{10}$ ${\rm g} \cdot {\rm cm}^{-3}$).
But, difference in such estimations is not principal for our analysis, so we will choose the density above given by Hamada and Salpeter for calculations.
We will show below that quantum corrections change results essentially also.



\begin{thebibliography}{00}
\bibitem{Cameron.1959b.AstrJ} 
  A.~G.~W.~Cameron,
\newblock
  \emph{Pycnonuclear reactions and nova explosions},
\newblock
  Astrophys. J. \textbf{130}, 916 (1959).

\bibitem{Zeldovich.1965.AstrJ}
  Ya.~B.~Zel'dovich and O.~H.~Guseynov,
\newblock
  \emph{Collapsed stars in binaries},
\newblock
  Astrophys. J. \textbf{144}, 840 (1965).

\bibitem{ShapiroTeukolsky.2004.book}
  S.~L.~Shapiro and S.~A.~Teukolsky,
\newblock
  \emph{Black Holes, White Dwarfs, and Neutron Stars: The Physics of Compact Objects}
\newblock
  (Wiley-VCH Verlag GmbH \& Co. KGaA, Weinheim, 2004), 645~pp.



\bibitem{Salpeter_VanHorn.1969.AstrJ} 
  E.~E.~Salpeter and H.~M.~Van Horn,
\newblock
  \emph{Nuclear reaction rates at high densities},
\newblock
  Astrophys. J. \textbf{155}, 183 (1969).

\bibitem{Schramm.1990.AstrJ}
  S. Schramm and S. E. Koonin, Astrophys. J. \textbf{365}, 296 (1990);
\newblock
  erratum: \textbf{377}, 343 (1991).

\bibitem{Haensel.1990.AstronAstrophys}
  P. Haensel and J. L. Zdunik, Astron. Astrophys. \textbf{229}, 117 (1990);
\newblock
  P. Haensel and J. L. Zdunik, Astron. Astrophys. \textbf{404}, L33 (2003).

\bibitem{Yakovlev.2006.PRC}
  D.~G.~Yakovlev, L.~R.~Gasques, M.~Beard, M.~Wiescher, and A.~V.~Afanasjev,
\newblock
  \emph{Fusion reactions in multicomponent dense mattter},
\newblock
  Phys. Rev. \textbf{C 74}, 035803 (2006);
\newblock
  arXiv:astro-ph/0608488.

\bibitem{Afanasjev.2010.ADNDT}
  M.~Beard, A.~V.~Afanasjev, L.~C.~Chamon, L.~R.~Gasques, M.~Wiescher, and D.~G.~Yakovlev,
\newblock
  \emph{Astrophysical S factors for fusion reactions involving \isotope{C}, \isotope{O}, \isotope{Ne}, and \isotope{Mg} isotopes},
\newblock
  At. Dat. Nucl. Dat. Tabl. \textbf{96}, 541--566 (2010);
\newblock
  arXiv:~1002.0741 [astro-ph.SR].


\bibitem{Singh.2019.NPA}
  V.~Singh, J.~Lahir, and D.~N.~Basu,
\newblock
  \emph{Theoretical exploration of $S$-factors for nuclear reactions of astrophysical importance},
\newblock
  Nucl. Phys. \textbf{A 987}, 260--273 (2019).


\bibitem{Afanasjev.2012.PRC}
  A.~V.~Afanasjev, M.~Beard, A.~I.~Chugunov, M.~Wiescher, and D.~G.~Yakovlev,
\newblock
  \emph{Large collection of astrophysical S-factors and its compact representation},
\newblock
  Phys. Rev. \textbf{C 85}, 054615 (2012);
\newblock
  arXiv:~1204.3174 [astro-ph.SR].

\bibitem{Maydanyuk.2011.JPS}
  S.~P.~Maydanyuk, S.~V.~Belchikov,
\newblock
  Journ. Phys. Stud. \textbf{14} (4), 40 (2011).

\bibitem{Maydanyuk.2015.NPA}
  S.~P.~Maydanyuk, P.-M.~Zhang, and S.~V.~Belchikov,
\newblock
  \emph{Quantum design using a multiple internal reflections method in a study of fusion processes in the capture of alpha-particles by nuclei},
\newblock
  Nucl. Phys. A \textbf{940}, 89--118 (2015);
\newblock
  arXiv:1504.00567.

\bibitem{Maydanyuk_Zhang_Zou.2017.PRC}
  S.~P.~Maydanyuk, P.-M.~Zhang, and L.-P.~Zou,
\newblock
  \emph{New quasibound states of the compound nucleus in $\alpha$-particle capture by the nucleus},
\newblock
  Phys. Rev. \textbf{C96}, 014602 (2017);
\newblock
  arXiv:1711.07012.


\bibitem{Landau.v3.1989}
  L.~D.~Landau and E.~M.~Lifshitz,
\newblock
  \textit{Kvantovaya Mehanika, kurs Teoreticheskoi Fiziki}
  (Quantum mechanics, course of Theoretical Physics),
\newblock
  Vol.~3 ({Nauka}, {Mockva}, 1989) p.~768 ---
  [in Russian; eng. variant: Oxford, Uk, Pergamon, 1982].

\bibitem{Eberhard.1979.PRL}
  K.~A.~Eberhard, Ch.~Appel, R.~Bangert, L.~Cleemann, J.~Eberth, and V.~Zobel,
\newblock
  \emph{Fusion cross sections for $\alpha + ^{40, 44}{\rm Ca}$ and the problem of anomalous large-angle scattering},
\newblock
  Phys. Rev. Lett. \textbf{43} (2), 107--110 (1979).

\bibitem{Gasques.2005.PRC}
  L.~R.~Gasques, A.~V.~Afanasjev, E.~F.~Aguilera, M.~Beard, L.~C.~Chamon, P.~Ring, M.~Wiescher, and D.~G.~Yakovlev,
\newblock
  \emph{Nuclear fusion in dense matter: Reaction rate and carbon burning},
\newblock
  Phys. Rev. \textbf{C 72}, 025806 (2005).


\bibitem{Maydanyuk.2011.JMP}
  S.~P.~Maydanyuk, S.~V.~Belchikov,
\newblock
  \emph{Problem of nuclear decay by proton emission in fully quantum consideration: Calculations of penetrability and role of boundary condition},
\newblock
  Journ. Mod. Phys. \textbf{2} (6), 572--585 (2011) [open access].

\bibitem{Maydanyuk.2000.UPJ}
  V.~S.~Olkhovsky, S.~P.~Maydanyuk
\newblock
  \textit{Method of multiple internal reflections in description of tunneling evolution through barriers},
\newblock
  Ukr. Phys. Journ. \textbf{45} (10), 1262--1269 (2000);
\newblock
  nucl-th/0406035.

\bibitem{Maydanyuk.2002.JPS}
  S.~P.~Maydanyuk, V.~S.~Olkhovsky, A.~K.~Zaichenko,
\newblock
  \textit{The method of multiple internal reflections in description of tunneling evolution of nonrelativistic particles and photons},
\newblock
  Journ. Phys. Stud. \textbf{6} (1), 1--16 (2002),
\newblock
  nucl-th/0407108.

\bibitem{Maydanyuk.2006.FPL}
  F.~Cardone, S.~P.~Maidanyuk, R.~Mignani, V.~S.~Olkhovsky,
\newblock
  \textit{Multiple internal reflections during particle and photon tunneling},
\newblock
  Found. Phys. Lett. \textbf{19} (5), 441--457 (2006).


\bibitem{Afanasjev.2007.PRC}
  L.~R.~Gasques, A.~V.~Afanasjev, M.~Beard, J.~Lubian, T.~Neff et al.,
\newblock
  \emph{Sao Paulo potential as a tool for calculating S factors of fusion reactions in dense stellar matter},
\newblock
  Phys. Rev. \textbf{C 76}, 045802 (2007).

\bibitem{Denisov.2015.PRC}
  V.~Yu.~Denisov,
\newblock
  \emph{Nucleus-nucleus potential with shell-correction contribution},
\newblock
  Phys. Rev. \textbf{C 91}, 024603 (2015).

\bibitem{Afanasjev.2006.PRC}
  L.~R.~Gasques, A.~V.~Afanasjev, E.~F.~Aguilera, M.~Beard, L.~C.~Chamon et al.,
\newblock
  \emph{Nuclear fusion in dense matter: Reaction rate and carbon burning},
\newblock
  Phys. Rev. \textbf{C 72}, 025806 (2006);
\newblock
  arXiv:~astro-ph/0506386.



\bibitem{Maydanyuk.2010.PRC}
  S.~P.~Maydanyuk, V.~S.~Olkhovsky, G.~Mandaglio, M.~Manganaro, G.~Fazio, and G.~Giardina,
\newblock
   \emph{Bremsstrahlung emission of high energy accompanying spontaneous of \isotope[252]{Cf}},
\newblock
  Phys. Rev. \textbf{C82}, 014602 (2010).

\bibitem{Maydanyuk.2010.IJMPE}
  S.~P.~Maydanyuk, V.~S.~Olkhovsky, G.~Mandaglio, M.~Manganaro, G.~Fazio, and G.~Giardina,
\newblock
   \emph{Bremsstrahlung emission accompanying decays and spontaneous fission of heavy nuclei},
\newblock
  Int. J. Mod. Phys. \textbf{E 19}, 1189--1196 (2010).

\bibitem{Maydanyuk.2011.JPCS}
  S.~P.~Maydanyuk, V.~S.~Olkhovsky, G.~Mandaglio, M.~Manganaro, G.~Fazio, and G.~Giardina,
\newblock
   \emph{Bremsstrahlung emission of photons accompanying ternary fission of \isotope[252]{Cf}},
\newblock
\newblock
  Journ. Phys.: Conf. Ser. \textbf{282}, 012016 (2011).






\bibitem{Hamada_Salpeter.1961.AstrJ}
  T.~Hamada and E.~E.~Salpeter,
\newblock
  \emph{Models for zero-temperature stars},
\newblock
  Astrophys. J. \textbf{134}, 683 (1961).

\bibitem{Maydanyuk.2011.EPJP}
  S.~P.~Maydanyuk,
\newblock
  \emph{Resonant structure of the early-universe space-time},
\newblock
  Europ. Phys. J. Plus \textbf{126}, 76 (2011),
\newblock
  arXiv:1005.5447.


\bibitem{Kelley.2017.NPA}
  J.~H.~Kelley, J.~E.~Purcell, and C.~G.~Sheu,
\newblock
  \emph{Energy levels of light nuclei $A = 12$},
\newblock
  Nucl. Phys. \textbf{A 968}, 71-253 (2017).


\bibitem{Fang.2017.PRC}
  X.~Fang, W.~P.~Tan, M.~Beard, R.~J.~deBoer, G.~Gilardy, et al., 
\newblock
  \emph{Experimental measurement of $\isotope[12]{C} + \isotope[16]{O}$ fusion at stellar energies},
\newblock
  Phys. Rev. \textbf{C 96}, 045804 (2017).


\end{thebibliography}
\end{document}